\input{amssymb.sty}
\documentstyle[aps,pre,floats,epsf,eqsecnum,epsfig]{revtex}
\def\Tc{T_{\mbox{\scriptsize c}}}
\def\rhoc{\rho_{\mbox{\scriptsize c}}}
\def\pc{p_{\mbox{\scriptsize c}}}
\def\muc{\mu_{\mbox{\scriptsize c}}}
\begin{document}

\twocolumn[\hsize\textwidth\columnwidth\hsize\csname@twocolumnfalse\endcsname

\title{Asymmetric Fluid Criticality II: Finite-Size Scaling for Simulations}

\author{Young C. Kim and Michael E. Fisher}
\address{Institute for Physical Science and Technology, University of Maryland, College Park, Maryland 20742}
\date{\today}

\maketitle
%\vspace{-.1in}
\begin{abstract}
The vapor-liquid critical behavior of intrinsically asymmetric fluids is studied in finite systems of linear dimensions, $L$, focusing on periodic boundary conditions, as appropriate for simulations. The recently propounded ``complete'' thermodynamic $(L\rightarrow\infty)$ scaling theory incorporating pressure mixing in the scaling fields as well as corrections to scaling $\mbox{[arXiv:cond-mat/0212145]}$, is extended to finite $L$, initially in a grand canonical representation. The theory allows for a Yang-Yang anomaly in which, when $L\rightarrow\infty$, the second temperature derivative, $(d^{2}\mu_{\sigma}/dT^{2})$, of the chemical potential along the phase boundary, $\mu_{\sigma}(T)$, diverges when $T\rightarrow\Tc -$. The finite-size behavior of various special {\em critical loci} in the temperature-density or $(T,\rho)$ plane, in particular, the $k$-inflection susceptibility loci and the $Q$-maximal loci --- derived from $Q_{L}(T,\langle\rho\rangle_{L}) \equiv \langle m^{2}\rangle^{2}_{L}/\langle m^{4}\rangle_{L}$ where $m \equiv \rho - \langle\rho\rangle_{L}$ --- is carefully elucidated and shown to be of value in estimating $\Tc$ and $\rhoc$. Concrete illustrations are presented for the hard-core square-well fluid and for the restricted primitive model electrolyte including an estimate of the correlation exponent $\nu$ that confirms Ising-type character. The treatment is extended to the canonical representation where further complications appear.

\vspace{0.4in}
\end{abstract}

]
%\pagebreak
%\tableofcontents
\section{Introduction and Overview}
\label{sec1}

True phase transitions arise in statistical mechanics only in the thermodynamic limit in which the volume of a system, $V \equiv L^{d}$ (in $d$ dimensions), and the number of particles in the system, $N$, go to infinity, while the density $\rho=N/V$ remains finite. In this limit, to be denoted for brevity simply by $L\rightarrow\infty$, the free energy and other quantities may exhibit singularities at a phase boundary or critical point as functions of the temperature or other thermodynamic fields. However, for finite systems as, in particular, realized in computer simulations, the free energy becomes analytic everywhere in the temperature and in other fields such as the chemical potential, $\mu$, and the pressure, $p$. Thus thermodynamic quantities that vary discontinuously or diverge in the thermodynamic limit become rounded when $L$ is finite.

Computer simulations have been useful in quantifying and gaining insights into phase transitions in various systems. Nevertheless, to obtain precise, sharp results from simulations --- inevitably performed on finite-systems --- one must perform appropriate extrapolations on the size, $L$, of the simulation ``box.'' Crucial questions then arise: How should one best estimate critical points from the finite-size data? And, especially: How can one reliably ascertain the critical universality class of particular model systems?

To study the statistical mechanics of finite systems, one must at the start address two basic issues, namely, the overall geometry of the system and the specific nature of the boundary conditions. Here we will have in mind general $d$-dimensional systems with periodic boundary conditions imposed on ``rectangular'' boxes of dimensions $L_{1}\times L_{2}\times\cdots\times L_{d}=V=L^{d}$ in which the ratios $L_{k}/L$ remains fixed (typically at $1$) when $L\rightarrow\infty$. Of course, this geometry combined with periodic boundary conditions has been used extensively in computer simulations for studies of the bulk properties of fluids.

In the case of critical phenomena in systems with a well defined {\em axis of symmetry} in some thermodynamic plane, notably model magnetic materials and analogous lattice gases \cite{ref1}, in which the critical density is trivially known and the variation with $(T-\Tc)$ is of primary interest, the long-established theory of finite-size scaling \cite{ref2,ref3} and its subsequent developments \cite{ref4,ref5,ref6}, has provided effective answers to many questions of how to extrapolate data for finite systems. However, two new issues that demand further consideration have recently come to the fore. These are, first, the desire to obtain precise, {\em unbiased} answers for the universal critical behavior of ``complex'' and, especially, {\em asymmetric} fluid systems --- in which, in particular, {\em both} the critical temperature $\Tc$, {\em and} the critical density $\rhoc$, must be accurately estimated \cite{ref7} --- and, second, the realization that the existence of a so-called Yang-Yang anomaly \cite{ref8,ref9} --- in which the chemical potential $\mu_{\sigma}(T)$ on the vapor-liquid phase boundary exhibits a {\em divergent curvature} when $T\rightarrow\Tc -$ --- requires a significant elaboration \cite{ref8,ref10,ref11} of earlier formulations of bulk, thermodynamic scaling for fluids \cite{ref12,ref13}.

The appropriately extended, ``complete'' scaling formulation for bulk properties (i.e., in the thermodynamic limit) that is needed to encompass a Yang-Yang anomaly \cite{ref8} has recently been carefully expounded and investigated in some detail: first in Part I of this article \cite{ref10}, to be denoted here as {\bf I}, and, more fully, in the thesis \cite{ref11} of the first author, which will be referred to here as {\bf K}. It proves necessary to ``mix'' the pressure, $p$, into the linear (and nonlinear) scaling fields \cite{ref8}. To be explicit, let us, following {\bf I}, introduce the dimensionless deviations from the (bulk) critical point $(\pc,\Tc,\muc)$ via
 \begin{equation}
  \check{p} = \frac{p-\pc}{\rhoc k_{\mbox{\scriptsize B}}\Tc}, \hspace{0.1in} t \equiv \frac{T-\Tc}{\Tc}, \hspace{0.1in} \check{\mu} = \frac{\mu - \muc}{k_{\mbox{\scriptsize B}}\Tc}.  \label{eq1.1}
 \end{equation}
Then the three relevant scaling fields for a single-component fluid must, in general, take the forms
 \begin{eqnarray}
   \tilde{p} & = & \check{p} - k_{0}t - l_{0}\check{\mu},  \label{eq1.2} \\
   \tilde{t} & = & t - l_{1}\check{\mu} - j_{1}\check{p},  \label{eq1.3} \\
   \tilde{h} & = & \check{\mu} - k_{1}t - j_{2}\check{p},  \label{eq1.4}
 \end{eqnarray}
in which the quadratic and higher order terms have been dropped (see {\bf I}). The crucial new feature (going beyond the previously accepted analyses: see \cite{ref12,ref13}) is the presence of the, in general, nonzero dimensionless {\em pressure-mixing coefficients} $j_{1}$ and $j_{2}\,$: when these vanish the earlier formulations are satisfactory.

In terms of the (nonlinear) scaling fields the general scaling hypothesis of {\bf I} asserts that the thermodynamics near criticality can be described, at least asymptotically, by
 \begin{equation}
  \Psi (\lambda^{2-\alpha}\tilde{p},\lambda\tilde{t},\lambda^{\Delta}\tilde{h};\lambda^{-\theta_{4}}u_{4},\lambda^{-\theta_{5}}u_{5},\cdots) = 0,   \label{eq1.5}
 \end{equation}
where $\lambda$ is a free, positive scaling parameter. The exponents $\alpha$ (for the specific heat) and $\Delta$ (for the ordering field $\tilde{h}$), are related to the other standard critical exponents via
 \begin{equation}
  \Delta = 2-\alpha-\beta = \beta + \gamma = \beta\delta,   \label{eq1.6}
 \end{equation}
while $\theta_{4}\equiv \theta$ and $\theta_{5}$ are the {\em positive} leading even and odd correction-to-scaling exponents for the corresponding {\em irrelevant} scaling fields, $u_{4}(p,T,\mu)$ and $u_{5}(p,T,\mu)$. One then discovers [8,{\bf I},{\bf K}] that the scaling form (\ref{eq1.5}) implies: (a) the existence of a Yang-Yang anomaly in which $(d^{2}\mu_{\sigma}/dT^{2})$ diverges as $\sim j_{2}/|t|^{\alpha}$ when $t\rightarrow 0$ and (b) a leading singular term varying as $\sim j_{2}|t|^{2\beta}$ in the coexistence curve diameter, that dominates the previously known term $\sim (l_{1}+j_{1})|t|^{1-\alpha}$ since, e.g., $\beta = 0.32_{6}$ and $\alpha=0.10_{9}$ for $d=3$ Ising-type criticality. Further new, singular terms of similar character appear in other thermodynamic properties: see {\bf I}.

The task addressed here, in Sec.\ II, is to systematically extend the general formulation for bulk scaling, as embodied in (\ref{eq1.1})-(\ref{eq1.5}), to finite systems characterized by a (single) finite length scale, $L$. According to the general principles of finite-size scaling, by which all lengths should, in the critical region, be scaled by the correlation length, $\xi(T) \sim 1/|t|^{\nu}$, we may anticipate that, in effect, the scaling parameter $\lambda$ in (\ref{eq1.5}) may, in a grand canonical setting, be replaced by $L^{1/\nu}$. Let us also note that when, as for real fluids, hyperscaling is valid (see Sec.\ II.A) we have
 \begin{equation}
   d\nu = 2-\alpha.   \label{eq1.6a}
 \end{equation}

It has, however, been pointed out \cite{ref14} that the scaling fields, $\tilde{p}$, $\tilde{t}$, $\tilde{h}$, $\cdots$, themselves may, in a finite system, gain an explicit dependence on the size $L$. Thus finite-size effects in a system confined by hard walls might well be dominated by $1/L$ contributions \cite{ref5,ref15}. This issue, which is by no means definitively settled in general, is considered briefly in Sec.\ II with the conclusion that for the case of periodic boundary conditions, which is our main concern here, one should anticipate additive terms in (\ref{eq1.2})-(\ref{eq1.4}); specifically, then, we will [setting $l_{0}\equiv 1$; see {\bf I}(3.22)] adopt the scaling field
 \begin{equation}
  \tilde{p}(p,T,\mu;L) = \check{p}-k_{0}t - \check{\mu} - s_{0}/L^{\bar{d}} + \cdots,   \label{eq1.7}
 \end{equation}
and likewise, with new coefficients $s_{1}$ and $s_{2}$, for $\tilde{t}(p,T,\mu;L)$ and $\tilde{h}(p,T,\mu;L)$, with $\bar{d}\geq 2$. (Note that the coefficients $s_{0}$, $s_{1}$, and $s_{2}$ carry dimensions of $L^{\bar{d}}$.) Fortunately, it then transpires that these $L$-dependent contributions do {\em not} enter the leading behavior of the quantities of principal interest, such as the $k$- and $Q$-loci in the $(T,\rho)$ plane: see below.

Specific predictions for the finite-size variation of basic densities and susceptibilities are presented in Sec.\ II.C. The variation with $L$ of the chemical potential, $\mu$, {\em at} the bulk critical temperature {\em and} density is examined in Sec.\ II.D: the answer provides a route to uncovering the presence of $L$-dependent terms in the scaling fields as in (\ref{eq1.7}).

Now, as mentioned, an important application of finite-size scaling theory is to analyze numerical data obtained from simulations on finite systems, and, thereby, to gain knowledge of the critical properties of the bulk system. Major efforts have been devoted to estimating critical parameters such as $(\Tc,\rhoc)$, and to confirming universality classes. As regards the estimation of $\Tc$ and $\rhoc$, most studies have focused on calculating the coexistence curve in the $(\rho,T)$ plane and then fitting the data with some suitably chosen formula in which $\Tc$ and $\rhoc$ appear.

However, simulations of a system in its two-phase region may require prohibitively long times or special, more elaborate computational techniques to equilibrate the two coexisting phases owing to the free-energy barrier that grows rapidly as $T$ decreases and $L$ increases. Moreover, since the correlation length, $\xi(T,\rho)$, becomes large and eventually diverges when the critical region is approached, finite-size effects smear out the vapor and liquid states near $\Tc$ and blur their distinction thereby seriously hampering the reliable determination of the coexistence curve. Finally, field mixing (even in the absence of pressure mixing) distorts the shape of the diameter, etc. Consequently, naively fitting coexistence curve data may yield quite poor values for $\Tc$ and $\rhoc$.

To meet these latter challenges, Bruce and Wilding some time ago \cite{ref16,ref17} proposed a rather convenient and effective finite-size scaling method for estimating $\Tc$ and $\rhoc$, that, in particular, incorporates $\check{\mu}$ and $t$ mixing into the scaling fields $\tilde{t}$ and $\tilde{h}$ (although pressure mixing is {\em not} included). Their method, which has proved quite popular, is based on the hypothesis that fluid criticality belongs to the Ising universality class (or, more generally, to some well studied universality class for which certain detailed critical properties are well established numerically). On that basis their method matches distribution functions of density {\em and} energy fluctuations observed in simulations to the (presumed available) limiting fixed point distributions as obtained {\em a priori} from simulations of simpler models (known to be of Ising or other character). In this way, following extrapolation on $L$, they estimate critical parameters. However, significant questions remain: What should be done when {\em a priori} knowledge of the (suspected or, possibly, quite new) critical behavior of the system of interest is {\em not} available? And; How should one proceed if the effects of pressure mixing may {\em not} be negligible? \cite{ref18}

In light of these serious issues, an important aim of our studies has been to develop {\em unbiased} finite-size scaling methods for estimating $\Tc$ and $\rhoc$ without the need for such strong assumptions and extensive {\em a priori} knowledge. For this purpose, as previously reported \cite{ref7,ref19,ref10}, various special loci have been introduced that, in the thermodynamic limit, spring from the critical point in the density-temperature or other thermodynamic plane. The bulk scaling behavior of these critical loci was derived within the complete, scaling theory in {\bf I} (and also studied there within classical mean-field theory). Among these loci, the $k$-loci --- defined via the points of isothermal maxima of $\chi^{(k)}=\chi/\rho^{k}$ in the $(\rho,T)$ plane, where $\chi\equiv \rho^{2}k_{\mbox{\scriptsize B}}TK_{T}$ is the isothermal susceptibility --- have already been used in simulations to estimate the critical points of the hard-core square-well (HCSW) fluid \cite{ref7}, and of the restricted primitive model (RPM) electrolyte \cite{ref19}. It is a goal of the present article to analyze the behavior of these $k$-loci in systems of {\em finite size}: explicit expressions for $\rho^{(k)}(T;L)$, the $k$-loci, in the $(\rho,T)$ plane are obtained in Sec.\ III. Not surprisingly, one finds that the density $\rho=\rho^{(k)}(\Tc;L)$ evaluated on a $k$-locus {\em at} $\Tc$ (where we suppose that $\Tc$ has been estimated reliably in some other way) approaches the critical density $\rhoc$ when $L\rightarrow\infty$: But in what manner?

We show in Sec.\ III.A that there is a leading deviation of magnitude $L^{-2\beta/\nu}$ followed by a term of order $L^{-(1-\alpha)/\nu}$: however, the amplitude of the leading contribution {\em vanishes} when $k$ takes an ``optimal'' value $k_{\mbox{\scriptsize opt}}=3{\cal R}_{\mu}$. In this result ${\cal R}_{\mu}$ is the (dimensionless) strength of the Yang-Yang anomaly as defined in Ref.\ \cite{ref8} and in {\bf I}.Sec.III.E. Extrapolating data for the densities $\rho^{(k)}(\Tc;L)$ to the thermodynamic limit can thus provide {\em unbiased} (bulk) estimates of the critical density. In Sec.\ III.B we re-apply this approach to the HCSW fluid using what we believe is an improved estimate for $\Tc$: see below. Our new estimate for $\rhoc$ agrees well, within the uncertainties, with the previous result \cite{ref7}. As indicated, this method for estimating $\rhoc$ has also been successfully applied to the RPM electrolyte \cite{ref19}.

Evidently, however, in locating $\rhoc$ by this route, one first needs a good estimate of $\Tc$. For fluids with relatively weak asymmetry like the hard-core square-well model, it was found \cite{ref7} that the extrema in density of the $k$-loci themselves provide fairly good estimators for $\Tc$ that may be extrapolated in $L$. However, the whole critical region of the RPM is extremely asymmetric, in part, so it seems, because of the remarkably low value, $\rhoc^{\ast}=\rhoc a^{3} \simeq 0.08$ \cite{ref19}, of the reduced critical density (where $a$ is the hard-core diameter). As a result, estimators for $\Tc$ based on the available $k$-loci prove rather misleading: indeed, the $k$-loci for ``near-optimal'' values of $k$ are observed to vary {\em nonmonotonically} in $\rho$ --- probably as a result of competition between the two leading contributions, $\Delta\rho \sim (k-k_{\mbox{\scriptsize opt}})/L^{2\beta/\nu}$ and $1/L^{(1-\alpha)/\nu}$, mentioned above. To overcome this serious obstacle to progress, Luijten, Fisher and Panagiotopoulos \cite{ref19} introduced the $Q$-loci which they defined by points of isothermal maxima in the $(\rho,T)$ plane of the inverse Binder parameter \cite{ref20}
 \begin{equation}
  Q_{L}(T;\langle\rho\rangle_{L}) = \frac{\langle m^{2}\rangle^{2}_{L}}{\langle m^{4}\rangle_{L}} \hspace{0.1in} \mbox{with} \hspace{0.1in} m = \rho-\langle\rho\rangle_{L},   \label{eq1.8}
 \end{equation}
where $\langle\cdot\rangle_{L}$ denotes a grand-canonical ensemble average in the finite system.

Now when $L\rightarrow\infty$ anywhere in the one-phase region one has $Q_{L}(T;\rho)\rightarrow\frac{1}{3}$ \cite{ref20}, where for brevity, we have replaced the argument $\langle\rho\rangle_{L}$ in $Q_{L}$ by $\rho$. On the other hand, {\em at} criticality, $Q_{L}(\Tc;\rhoc)$ approaches a universal value $Q_{\mbox{\scriptsize c}}$ that is close to 0.6326 for $(d=3)$-dimensional Ising systems in a cubic box with periodic boundary conditions \cite{ref21,ref22,ref22a}. For finite systems at fixed $T$ near criticality, however, one finds that $Q_{L}$ exhibits rounded maxima that serve to provide well-defined loci, $\rho_{Q}(T;L)$ \cite{ref19}. The behavior of these $Q$-loci for large $L$ is derived explicitly within the full finite-size scaling theory in Sec.\ IV.A. One might note that determining the $Q$-loci involves calculation of the fourth density moment and of its density derivative (i.e., the fifth moment) so that the analysis requires some care! By the same token, in order to obtain the $Q$-loci reliably via simulations, data of high quality are needed. As for the $k$-loci, one may define $Q^{(k)}$-loci by points of isothermal maxima in the $(\rho,T)$ plane of a modified $Q$ parameter, namely, $Q^{(k)}\equiv Q_{L}/\rho^{k}$. The behavior of these loci is presented in Sec.\ IV.B: we find that the density, $\rho_{Q}^{(k)}(\Tc;L)$, evaluated at $\Tc$ on these loci varies in leading order as $L^{-2\beta/\nu}$ with, as in the $k$-loci, a subsequent $L^{-(1-\alpha)/\nu}$ term. However, the amplitude of the leading contribution now vanishes when $k=-9{\cal R}_{\mu}$, in contrast to $k_{\mbox{\scriptsize opt}}=3{\cal R}_{\mu}$ for the $k$-loci; thus the ``optimal'' value of $k$ for the $Q^{(k)}$-loci has the opposite sign.

Following Binder's original approach for {\em symmetric} systems \cite{ref20}, Luijten {\em et al.} \cite{ref19} examined plots of
 \begin{equation}
  Q_{L}^{Q}(T) \equiv Q_{L}\mbox{\boldmath $($}T;\rho_{Q}(T;L)\mbox{\boldmath $)$},   \label{eq1.9}
 \end{equation}
i.e., $Q_{L}$ evaluated {\em on} the $Q$-loci $\rho_{Q}(T,L)$. For the RPM they observed that the successive self-intersections as $L$ increased, say $\Tc^{Q}(L)$, converged rather rapidly to a precisely defined value, $T_{\infty}$ --- which thus served as a good estimate for $\Tc$. At the same time, they surprisingly found that the values of $Q_{L}^{Q}$ at the intersection points approached a limit which could be identified as a (surprisingly precise) estimate of the universal value $Q_{\mbox{\scriptsize c}}$. Thereby they established convincingly that the RPM (at least within the $\zeta=5$ level of discretization they studied \cite{ref19}) belongs to the short-range Ising universality class --- despite the long-range Coulomb interactions in the model. We show here that the approach of the estimators, $\Tc^{Q}(L)$, derived from the $Q_{L}^{Q}(T)$ plots to $\Tc$ obeys a $1/L^{(1+\theta)/\nu}$ law, while the difference $Q_{L}^{Q}\mbox{\boldmath $($}\Tc(L)\mbox{\boldmath $)$}-Q_{\mbox{\scriptsize c}}$ varies as $L^{-\theta/\nu}$ followed by a $j_{2}^{2}L^{-2\beta/\nu}$ term (see Sec.\ V.B). Note that these results are {\em independent} of asymmetry or pressure mixing (in leading order). 

In Sec.\ V.A we develop the theory for this approach and apply it to re-estimate $\Tc$ for the HCSW model \cite{ref7}. The new estimate is about $0.06\%$ higher than the earlier value \cite{ref7}; but that leads to no significant changes in the main conclusions reached previously: in particular, as noted above, the previous estimate for $\rhoc$ remains unchanged (within the uncertainties).

On the other hand, in Sec.\ IV.C we consider the behavior of $Q_{L}(T;\langle\rho\rangle_{L})$ for large $L$ in the two-phase region beneath $\Tc$. (See also Rovere, Heermann and Binder \cite{ref23}.) We exhibit plots for the HCSW fluid and RPM that illustrate some striking features (and we correct a misleading expression given in \cite{ref7} for the behavior of $Q_{L}(T;\rho_{L})$ with $\rho_{L}=\langle\rho\rangle_{L}$ when $L\rightarrow\infty$ below $\Tc$). In Sec.\ IV.D we go on to discuss the explicit scaling form for two {\em minima} of $Q_{L}(T;\rho_{L})$ that, when $T<\Tc$, approach the two sides of the coexistence curve rather rapidly as $L\rightarrow\infty$: see Figs.\ 8 and 9, below. It turns out that these considerations lead to a novel and apparently very effective and systematic method of estimating the limiting coexistence curve width and diameter, namely, 
 \begin{eqnarray}
  \Delta\rho_{\infty}(T) & \equiv & \rho_{+}(T)-\rho_{-}(T),   \label{eq1.10} \\
  \rho_{d}(T) & = & \mbox{$\frac{1}{2}$}[\rho_{+}(T)+\rho_{-}(T)],  \label{eq1.11}
 \end{eqnarray}
where $\rho_{+}(T) \equiv \rho_{\mbox{\scriptsize liq}}(T)$ and $\rho_{-}(T) \equiv \rho_{\mbox{\scriptsize vap}}(T)$ denote the true, bulk liquid and vapor densities, respectively. This method, which yields precise results surprisingly close to $\Tc$, has been applied to the HCSW and RPM models; however, the details, which entail using the simulation data to generate a scaling function for the minima as $T\rightarrow\Tc -$, will be expounded elsewhere \cite{ref24}.

The universality class of a particular system can be identified or checked and confirmed by determining critical exponents, $\alpha$, $\beta$, etc. In Sec.\ V.C we analyze further a method for estimating the correlation-length exponent, $\nu$ \cite{ref7}. This method has been applied to the HCSW fluid \cite{ref7} and, more recently, reported for the RPM electrolyte \cite{ref19}. A thermodynamic quantity for a finite system, say $P_{L}(T)$, evaluated on some suitable locus, say $\rho=\rhoc$, may exhibit a maximum at $T=\Tc^{P}(L)$ which can be regarded as an effective finite-size critical temperature. According to finite-size scaling one expects $\Tc^{P}(L)$ to approach the true critical temperature $\Tc$ asymptotically as $L^{-1/\nu}$. We confirm that this conclusion survives pressure-mixing (for suitable loci) and, by way of an application, show that by examining a rather wide range of properties $P_{L}(T)$ for the RPM one can identify those for which the desired maxima approach $\Tc$ {\em from above}. This is important in practice because simulations above criticality are significantly less hampered by problems of full equilibration than those at or below $\Tc$ where two distinct putative phases coexist, and ``alternate'' in the simulation box. Consequently, sufficiently {\em precise} calculations of $\Tc^{P}(L)$ are relatively easy which, in turn, provides a suitable basis for robust extrapolation. In this way, we show that one can estimate the exponent $\nu$ fairly accurately. For the RPM electrolyte (at the $\zeta=5$ level of discretization) we find $\nu = 0.63 \pm 0.03$ \cite{ref19} which supports the conclusion that the model belongs to the $(d=3)$-dimensional Ising universality class \cite{ref19}.

Both for gaining insight into experiments, in which the density $\rho$ is most often a controlled variable, and, likewise, for simulations in which the particle number, $N$, is fixed, it is valuable to study the finite-size scaling behavior of near-critical fluids in a canonical or $(\rho,T)$ representation. The bulk canonical free energy density $f(\rho,T)=\lim_{L\rightarrow\infty} F_{N}(V,T)/V$, where $F_{N}(V,T)$ is the Helmholtz free energy, has a leading asymptotic scaling behavior near criticality of the form
 \begin{equation}
  f(\rho,T) \approx f_{0}(\rho,T) + A|t|^{-(2-\alpha)}X_{\pm}(m/|t|^{\beta}),   \label{eq1.12}
 \end{equation}
in which $f_{0}(\rho,T)$ is a smooth (generally  analytical) background part of the free energy while $m \equiv (\rho-\rhoc)/\rhoc$. However, this simple scaling form does not incorporate any mixing in the scaling fields. We may anticipate that upon incorporating the mixing of the scaling fields, the leading scaling behavior remains unchanged but with some modifications of the scaling variables $m$ and $t$. But what should be expected {\em precisely}? That may well affect the behavior of the corrections on various loci \cite{ref25}. And what scaling form should one obtain if, in particular, pressure-mixing is introduced? In Sec.\ VI we derive explicit canonical scaling forms from the complete scaling formulation in the grand canonical representation. This is carried out first for the thermodynamic limit: then our finite-size results are applied to obtain corresponding canonical expressions. In Sec.\ VI.B we discuss the definition of finite-size canonical critical points and elucidate their behavior as illustrated by results for the HCSW fluid and the RPM electrolyte \cite{ref7,ref19}.

Finally, Sec.\ VII summarizes the article briefly.

\section{ Full Finite-Size Scaling Formulation}
\label{sec2}
Here we extend to finite systems near bulk critical points the complete scaling theory that incorporates {\em pressure mixing} \cite{ref10}.

\subsection{Scaling functions and hyperuniversality}
\label{sec2.a}

To extend the bulk scaling ansatz (\ref{eq1.5}) to a finite $V$=$L^{d}$ system we first replace $\tilde{p}$, $\tilde{h}$, $\tilde{t}$; $u_{4}$, $u_{5}$, $\cdots$ by corresponding finite-size nonlinear scaling fields $\tilde{p}(p,T,\mu;L)$, $\cdots ;$ $\cdots,$ $u_{j}(p,T,\mu;L)$, $\cdots$ of the form (\ref{eq1.7}), etc, and choose an arbitrary fixed reference length, say $l_{\ast}$. Setting $\lambda = (L/l_{\ast})^{1/\nu}$ in (\ref{eq1.5}) then leads to the general hypothesis
 \begin{equation}
  \Psi \left( \tilde{p}\hspace{-0.035in}\left(\frac{L}{l_{\ast}}\right)^{\frac{\mbox{\scriptsize $2-\alpha$}}{\mbox{\scriptsize $\nu$}}}\hspace{-0.05in},\,\,\tilde{t}\hspace{-0.035in}\left(\frac{L}{l_{\ast}}\right)^{\frac{\mbox{\scriptsize $1$}}{\mbox{\scriptsize $\nu$}}}\hspace{-0.05in},\,\,\tilde{h}\hspace{-0.035in}\left(\frac{L}{l_{\ast}}\right)^{\frac{\mbox{\scriptsize $\Delta$}}{\mbox{\scriptsize $\nu$}}}\hspace{-0.05in};\, u_{4}\hspace{-0.035in}\left(\frac{l_{\ast}}{L}\right)^{\frac{\mbox{\scriptsize $\theta_{4}$}}{\mbox{\scriptsize $\nu$}}}\hspace{-0.05in},\,\,\cdots\right) = 0,   \label{eq2.1}
 \end{equation}
which we expect to be at least asymptotically valid for $L/l_{\ast}\rightarrow\infty$ as $\tilde{p}$, $\tilde{t}$ and $\tilde{h} \rightarrow 0$.

Let us now restrict attention to dimensionalities $d$ less than the upper critical dimensionality $d_{>} (=4$ for normal fluid criticality). Then the {\em hyperuniversality exponent relation}, supported by renormalization group (RG) theory (for a fixed point without dangerous irrelevant variables \cite{ref27}) dictates $(2-\alpha)/\nu = d$ [see (\ref{eq1.6a})] and we may solve (\ref{eq2.1}) for $\tilde{p}$ to obtain
 \begin{equation}
  \rhoc \tilde{p}(p,T,\mu;L) = L^{-d}Y(x_{L},y_{L}; y_{L4}, \cdots),  \label{eq2.2}
 \end{equation}
where we have introduced the dimensionless scaled variables
 \begin{eqnarray}
  x_{L} & = & D_{L}\tilde{t}L^{1/\nu}, \hspace{0.1in} y_{L}=U_{L}\tilde{h}L^{\Delta/\nu}, \hspace{0.1in} y_{Lk} = U_{Lk}L^{-\theta_{k}/\nu} \nonumber \\
   &  & \hspace{1.4in} (k=4,5,\cdots).  \label{eq2.3}
 \end{eqnarray}
Here $D_{L}$, $U_{L}$, and $U_{Lk}\propto u_{k}$ are {\em nonuniversal metrical factors}, of dimensions $l_{\ast}^{-1/\nu}$, $l_{\ast}^{-\Delta/\nu}$, $l_{\ast}^{\theta_{k}/\nu}$, respectively, which depend on the system under study.

By construction (note the factor $\rhoc >0$) the scaling function $Y(x,y;y_{4},\cdots)$ is dimensionless \cite{ref28}. However, the {\em hyperuniversality scaling hypothesis} \cite{ref29} (supported by various exact calculations \cite{ref29,ref30,ref31}, simulations \cite{ref32} and RG theory \cite{ref21}) tells us that $Y(x,y;y_{4},\cdots)$ is a {\em universal function} of its (appropriately normalized) arguments. Note, however, that $Y(x,y;y_{4},\cdots)$ {\em must} depend on the {\em geometry} of the finite system {\em and} on the {\em boundary conditions} imposed; but it will not depend on any microscopic details beyond those that determine the bulk universality class of the relevant critical point. Furthermore, $Y$ must be even under change of sign of the {\em odd scaling variables}, $y \Longleftrightarrow -y$, $y_{5}\Longleftrightarrow -y_{5}$, $\cdots$.

The bulk limit may be obtained formally by setting $L=1/|D_{L}\tilde{t}|^{\nu}$ and letting $L\rightarrow\infty$ (when it drops out of the nonlinear scaling fields $\tilde{p}$, $\tilde{t}$, $\cdots$). This yields the scaling form {\bf I}(2.3), namely,
 \begin{equation}
   \tilde{p} = Q|\tilde{t}|^{2-\alpha}W_{\pm}(y;y_{4},y_{5},\cdots),  \label{eq2.4}
 \end{equation}
with the identification $Q=|D_{L}|^{2-\alpha}/\rhoc$, which is, thus, a dimensionless {\em non}universal amplitude, while 
 \begin{equation}
   W_{\pm}(y;y_{4},\cdots) = Y(\pm 1,y;y_{4},\cdots),  \label{eq2.5}
 \end{equation}
is universal with the amplitudes in {\bf I}(2.1) and {\bf I}(2.2) related by $U$=$U_{L}/|D_{L}|^{\Delta}$, $U_{k}$=$U_{Lk}|D_{L}|^{\theta_{k}}\,\, (k=4,5,\cdots)$.

In contrast to the bulk scaling function, the finite-size function, $Y(x,y;y_{4},\cdots)$ must be analytic in the vicinity of the origin since all critical singularities will be rounded in a finite system. Following {\bf I} we may thus expand for large $L$ in powers of the irrelevant variables to obtain
 \begin{equation}
  Y(x_{L},y_{L};\cdots) = Y^{0}(x_{L},y_{L}) + \sum_{\mbox{\boldmath\scriptsize $\kappa$}} Y^{\mbox{\boldmath\scriptsize $\kappa$}}(x_{L},y_{L})y^{[\mbox{\boldmath\scriptsize $\kappa$}]},  \label{eq2.6}
 \end{equation}
where, as in {\bf I}, the multi-index, {\boldmath $\kappa$}, is defined by {\boldmath $\kappa$} $=$ (4), (5), $\cdots$, (4,4), (4,5), $\cdots$, (4,4,4), $\cdots$, while $y^{[i,j,\cdots,n]}$ means $y_{Li}y_{Lj}\cdots y_{Ln}$. The underlying symmetry of the scaling function, $Y(x_{L},y_{L};\cdots)$, that is evidenced by exact results and RG theory, then requires
 \begin{equation}
  Y^{\mbox{\boldmath\scriptsize $\kappa$}}(x_{L},-y_{L}) = \pm Y^{\mbox{\boldmath\scriptsize $\kappa$}}(x_{L},y_{L}),   \label{eq2.7}
 \end{equation}
for {\boldmath $\kappa$} even or odd in the sense of {\bf I}(2.7). Thence we have the expansions
 \begin{eqnarray}
   Y^{\mbox{\boldmath\scriptsize $\kappa$}}(x_{L},y_{L}) & = & Y_{00}^{\mbox{\boldmath\scriptsize $\kappa$}} + Y_{10}^{\mbox{\boldmath\scriptsize $\kappa$}}x_{L} + Y_{20}^{\mbox{\boldmath\scriptsize $\kappa$}}x_{L}^{2} + Y_{02}^{\mbox{\boldmath\scriptsize $\kappa$}}y_{L}^{2} + \cdots,  \nonumber \\
  & = & y_{L}( Y_{01}^{\mbox{\boldmath\scriptsize $\kappa$}} + Y_{11}^{\mbox{\boldmath\scriptsize $\kappa$}}x_{L} + Y_{21}^{\mbox{\boldmath\scriptsize $\kappa$}}x_{L}^{2} + Y_{03}^{\mbox{\boldmath\scriptsize $\kappa$}}y_{L}^{2} + \cdots), \nonumber \\  \label{eq2.8}
 \end{eqnarray}
for {\boldmath $\kappa$}, even and odd, respectively, where the expansion coefficients $Y_{ij}^{\mbox{\boldmath\scriptsize $\kappa$}}$ are universal numbers.

For our present purposes the leading approximation
 \begin{equation}
  Y \approx Y^{0}(x_{L},y_{L}) + y_{L4}^{\mbox{\scriptsize c}}Y^{(4)}(x_{L},y_{L}) + y_{L5}^{\mbox{\scriptsize c}}Y^{(5)}(x_{L},y_{L}),  \label{eq2.9}
 \end{equation}
in which $U_{L4}$ and $U_{L5}$ in the definitions of $y_{L4}$ and $y_{L5}$ have been replaced by their critical-point values, will amply suffice.

\subsection{Finite-size corrections to the scaling fields}
\label{sec2.b}

In this section we discuss in a little more detail the question of finite-size corrections to the scaling fields that was touched on in the Introduction. This issue seems to have been first raised in Ref.\ \cite{ref14} but to have escaped much more extensive or systematic discussion. Here we consider only a $d$-dimensional hypercube with periodic boundary conditions.

A field-theoretic RG approach to finite-size scaling was initiated by Br\'{e}zin \cite{ref33}. Later, with Zinn-Justin \cite{ref21} systematic calculations of the scaling functions were presented using both $d=4-\epsilon$ and $d=2+\epsilon$ expansions. In particular, the shift of $\Tc$ that enters the scaling variable, $\tilde{t}$, of the universal scaling functions was computed: see Ref.\ \cite{ref21} Eqns.\ (3.20) and (3.32). Indeed, $\tilde{t}$ as calculated in Eq.\ (3.21) of Ref.\ \cite{ref21} contains finite-size corrections that, in leading order, vary as $L^{-2}$. A similar form for $\tilde{t}$ was obtained by Korutcheva and Tonchev \cite{ref34} for a finite system with long-range interactions decaying as $1/r^{d+2-2\sigma}$, $\sigma \rightarrow 0+$. Recently, Chen and Dohm \cite{ref35} calculated the finite-size free-energy density of an $O(n)$ $\varphi^{4}$ field theory confined in a hypercube with periodic boundary conditions: they used a {\em sharp} cutoff in {\bf k} space and obtained a nonuniversal $L^{-2}$ contribution that dominated a universal scaling part that varied as $L^{-d}$.

On the other hand, Jasnow and coworkers \cite{ref31,ref36} concluded via RG theory that the system size $L$ does {\em not} enter in the formation of the scaling fields: see, especially Ref.\ \cite{ref31} Sec.\ III. Likewise Zinn-Justin [6, page 778] argues that: ``The crucial observation which explains finite-size scaling is that the renormalization theory which leads to RG equations is completely {\em insensitive to finite size effects} since renormalizations are entirely due to {\em short distance singularities}. As a consequence RG equations are not modified. $\cdots$''. Nevertheless, in our assessment it remains uncertain whether or not, even in the simplest case of periodic boundary conditions, the system size affects the scaling fields. While further careful analyses may settle the issue convincingly, we feel justified in allowing for an $L^{-\bar{d}}$ leading contribution in all the scaling fields --- as embodied in (\ref{eq1.7}); however, it seems safe to assume that $\bar{d}\geq 2$. As mentioned in the Introduction, we then find in most cases that these corrections are {\em less} important, when $L$ becomes large, than those arising from field mixing and the leading irrelevant variables.

\subsection{Some basic thermodynamic properties}
\label{sec2.c}

The generalized number and entropy ``scaling'' densities, $\tilde{\rho}$ and $\tilde{s}$, introduced in {\bf I} play a significant role also in analyzing finite systems: they are defined by
 \begin{equation}
  \tilde{\rho} \equiv (\partial\tilde{p}/\partial\tilde{h})_{\tilde{t}}, \hspace{0.3in} \tilde{s} \equiv (\partial\tilde{p}/\partial\tilde{t})_{\tilde{h}}.   \label{eq2.10}
 \end{equation}
From (\ref{eq2.2}) and (\ref{eq2.3}), we obtain \cite{ref28}
 \begin{equation}
  \rhoc \tilde{\rho} = U_{L}L^{-\beta/\nu} (\partial_{y} Y), \hspace{0.05in} \rhoc \tilde{s} = D_{L}L^{-(1-\alpha)/\nu}(\partial_{x}Y),  \label{eq2.11}
 \end{equation}
where, here and below, we adopt the notations $(\partial_{x}Y) \equiv (\partial Y/\partial x_{L})_{y_{L}}$, etc.

Now recall the definitions {\bf I}(2.14) of the ``true'' reduced number and entropy densities, namely,
  \begin{equation}
    \check{\rho} \equiv \frac{\rho}{\rhoc} = \left(\frac{\partial\check{p}}{\partial\check{\mu}}\right)_{t}, \hspace{0.1in} \check{s} = \frac{\cal S}{\rhoc k_{\mbox{\scriptsize B}}} = \left(\frac{\partial\check{p}}{\partial t}\right)_{\mu}.  \label{eq2.12}
  \end{equation}
Following {\bf I}(2.16)-(2.19) these may be expressed in terms of the generalized, scaling densities. Thus we find
  \begin{eqnarray}
   \check{\rho} & = & l_{0} + (2q_{0}+l_{0}n_{0})\check{\mu} + (n_{0}+2l_{0}m_{0})\check{p} + (v_{0}+l_{0}n_{3})t \nonumber \\
  &  & +\: (1-j_{2}l_{0})\tilde{\rho} - (l_{1}+j_{1}l_{0})\tilde{s}+j_{2}(j_{2}l_{0}-1)\tilde{\rho}^{2} \nonumber \\
  &  & +\: O(\tilde{\rho}\tilde{s},\tilde{s}^{2}),   \label{eq2.13}
  \end{eqnarray}
where $q_{0}$, $n_{0}$, $m_{0}$, $v_{0}$, $n_{3}$, etc. are the quadratic mixing coefficients entering the full nonlinear scaling fields: see {\bf I}(1.4)-(1.6); in addition, one discovers that the finite-size $L^{-\bar{d}}$ correction terms in the scaling fields --- see (\ref{eq1.7}) --- enter only with the quadratic mixing coefficients. Likewise we obtain
  \begin{eqnarray}
    \check{s} & = & k_{0} + (v_{0}+k_{0}n_{0})\check{\mu} + (n_{3}+2k_{0}m_{0})\check{p} + (2r_{0}+k_{0}n_{3})t \nonumber \\
  &  & -\: (k_{1}+j_{2}k_{0})\tilde{\rho} + (1-j_{1}k_{0})\tilde{s} + O(\tilde{\rho}^{2},\tilde{\rho}\tilde{s},\tilde{s}^{2}),  \label{eq2.14}
  \end{eqnarray}
where, again, we have retained only the leading terms needed later: further terms are given in {\bf K}(4.29)-(4.30).

Similarly, the generalized susceptibilities defined in {\bf I}(2.28) are useful here: one finds
  \begin{equation}
    \tilde{\chi}_{hh} \equiv (\partial^{2}\tilde{p}/\partial\tilde{h}^{2})_{\tilde{t}} = U_{L}^{2}L^{\gamma/\nu}(\partial_{y}^{2}Y)/\rhoc,   \label{eq2.15}
  \end{equation}
and likewise for $\tilde{\chi}_{ht}$ and $\tilde{\chi}_{tt}$. The basic number fluctuation or reduced susceptibility $\check{\chi}_{NN} = (\partial^{2}\check{p}/\partial\check{\mu}^{2})_{t}$ can then --- see {\bf I}(2.29) and {\bf K}(4.33) and Appendix F --- be expressed as
  \begin{eqnarray}
   \rhoc\check{\chi}_{NN} & = & e_{1}^{2}U_{L}^{2}L^{\gamma/\nu}(\partial_{y}^{2}Y) \nonumber \\
   &  & -\: 3j_{2}e_{1}U_{L}^{3}L^{(\gamma-\beta)/\nu}(\partial_{y}^{2}Y)(\partial_{y}Y)/\rhoc  \nonumber \\
  &  & -\: 2e_{1}e_{3}U_L D_L L^{(1-\beta)/\nu}(\partial_x \partial_y Y) + \cdots,  \label{eq2.16}
  \end{eqnarray}
where only the leading terms have been displayed while the constants are
  \begin{equation}
   e_{1} = 1 - j_{2},  \hspace{0.1in}  e_{3} = l_{1} + j_{1},  \hspace{0.1in} (l_{0}=1);  \label{eq2.17}
  \end{equation}
see {\bf I}(2.30) and {\bf I}(3.22). This result is needed to study the $k$-loci in finite systems: see Sec.\ III.A. The $Q$-loci, taken up in Sec.\ IV.A, demand the higher-order analogs.

\subsection{Chemical potential at {\boldmath $(T_{\mbox{\scriptsize\bf c}},\rho_{\mbox{\scriptsize\bf c}})$}}
\label{sec2.d}

Before turning to the various critical loci and their finite-size behavior, we address a rather special question which turns out to be interesting since its answer, as mentioned in the Introduction, opens an opportunity to determine via precise simulations, the presence or absence of finite-size dependence in the scaling fields. In a finite grand canonical ensemble at temperature $T$ the chemical potential $\mu$ must be adjusted to achieve a specified density: but the resulting value will depend on $L$. Accordingly we ask: ``How does the finite-size chemical potential, say $\mu_{L}^{\mbox{\scriptsize c}} \equiv \mu_{L}(\Tc,\rhoc)$, needed to achieve the bulk critical density, $\rhoc$, at the critical temperature, $\Tc$, approach $\mu_{\infty}^{\mbox{\scriptsize c}} \equiv \mu_{\mbox{\scriptsize c}}$?''

To attack the problem we first determine the scaling fields at $T=\Tc$ and $\rho=\rhoc$, i.e., $t=0$ and $\check{\rho}=\check{\rhoc}=1$. Recalling that $l_{0}=1$ [{\bf I}(3.22)], the relation (2.13) for the density $\check{\rho}$ then yields
  \begin{equation}
   0 = (1-j_{2})\tilde{\rho} - (l_{1}+j_{1})\tilde{s} - j_{2}(1-j_{2})\tilde{\rho}^{2} + \cdots,  \label{eq2.18}
  \end{equation}
where we have neglected the ``background'' terms in $\check{\mu}$ and $\check{p}$ (arising from the quadratic mixing coefficients) and may check later that they yield only higher-order corrections. (Note that $\tilde{s}\sim L^{-(1-\alpha)/\nu}$ dominates $L^{-\bar{d}}$ since $\bar{d}\geq 2 > (1-\alpha)/\nu$.) By appealing to (\ref{eq2.11}) and the scaling function expansions (\ref{eq2.9}) and (\ref{eq2.8}) this can be re-expressed as
  \begin{eqnarray}
  &  &  2(1-j_{2})U_{L}[Y_{02}^{0}+Y_{02}^{(4)}y_{L4}^{c} + \cdots]y_{L} \nonumber \\
  &  & - (l_{1}+j_{1})D_{L}L^{(\beta -1+\alpha)/\nu}[Y_{10}^{0} + \cdots] \approx 0.  \label{eq2.19}
  \end{eqnarray}
From the definitions (\ref{eq2.3}) of $y_{L}$ and $y_{Lk}$ we thus find that when $\rho=\rhoc$ at $t=0$ the ordering field obeys
  \begin{equation}
   \tilde{h} \approx  a_{\mu}/L^{(1-\alpha+\gamma)/\nu} = a_{\mu}/L^{d+(\gamma-1)/\nu},   \label{eq2.20}
  \end{equation}
where the omitted correction factor includes $L^{-\theta_{4}/\nu}$ and $L^{-1/\nu + d-\bar{d}}$ as leading contributions, while
  \begin{equation}
   a_{\mu} = (l_{1}+j_{1})D_{L}Y_{10}^{0}/2(1-j_{2})U_{L}^{2}Y_{02}^{0}.   \label{eq2.21}
  \end{equation}
Note also that even in the absence of pressure mixing (i.e., $j_1 = j_2 = 0$) the contribution of $\mu$ to $\tilde{t}$, via $l_1 \neq 0$, ensures that $\tilde{h}$ does not vanish (as it would identically in a symmetric system); instead $\tilde{h}$ decays with a leading exponent $d + (\gamma-1)/\nu$ of value about 3.38 for the $d=3$ Ising universality class.

Finally, at $t=0$ the relation (\ref{eq1.4}) for $\tilde{\mu}$ with the added term $-s_{2}/L^{\bar{d}}$, and (\ref{eq1.7}), leads, in linear order, to
  \begin{eqnarray}
   \check{\mu} & = & \tilde{h} + j_2 \check{p} + s_2 /L^{\bar{d}} \nonumber \\
   & = & \tilde{h} + j_2 (\tilde{p} + \check{\mu} + s_{0}/L^{\bar{d}}) + s_2 /L^{\bar{d}}.   \label{eq2.22}
  \end{eqnarray}
On using (\ref{eq2.2}) for $\tilde{p}$ at $x_L \approx y_L \approx 0$ this may be solved to yield
  \begin{eqnarray}
   \check{\mu}_L^{\mbox{\scriptsize c}} & \equiv & [\mu(\Tc,\rhoc;L)-\mu_{\mbox{\scriptsize c}}]/k_{\mbox{\scriptsize B}}\Tc,   \nonumber \\
  & = & a_{L}/L^{\bar{d}} + a_{p}/L^{d} + a_{\mu}/(1-j_2)L^{d+(\gamma-1)/\nu} + \cdots, \nonumber \\   \label{eq2.23}
  \end{eqnarray}
where the new amplitudes are
  \begin{equation}
   a_L = (s_2 + j_2 s_0)/(1-j_2 ), \hspace{0.05in} a_p = j_2 Y_{00}^0 /\rhoc (1-j_2 ).  \label{eq2.24}
  \end{equation}
Evidently, if $\bar{d}<d$ and $j_2 s_0$ and $s_2$ do not both vanish, the dominant behavior arises from the $L$-dependence of the scaling fields. If pressure mixing is absent (or negligible) the last, most rapidly decaying term in (\ref{eq2.23}) will be controlling.

\section{Modified-susceptibility loci in finite systems}
\label{sec3}
\subsection{ Asymptotic expressions}
\label{sec3.1}

The $k$-modified-susceptibility loci or, for brevity, the $k$-loci are defined by the isothermal maxima of $\chi^{(k)}\equiv \chi/\rho^{k}$ and so satisfy {\bf I}(4.32), namely,
  \begin{equation}
   \check{\rho}(\partial\check{\chi}_{NN}/\partial\check{\mu})_{T} = k(\check{\chi}_{NN})^{2}.  \label{eq3.1}
  \end{equation}
We aim to solve this equation asymptotically near criticality, first, to obtain $\check{\mu}^{(k)}(t;L)$, i.e., the finite-size $k$-loci in the $(\mu,T)$ plane, then $\check{p}^{(k)}(t;L)$, and, finally, $\check{\rho}^{(k)}(t;L)$, the locus in the $(\rho,T)$ plane which is of most practical interest.

The required third order susceptibility, $\check{\chi}_{N^{3}} \equiv (\partial\check{\chi}_{NN}/\partial\check{\mu})_{T}$, can be obtained by differentiating (\ref{eq2.16}) with respect to $\check{\mu}$ at fixed $t$. This entails the derivatives
  \begin{eqnarray}
   (\partial x_{L}/\partial\check{\mu})_{T} & = & D_{L}L^{1/\nu}(-l_{1}-j_{1}\check{\rho}+\cdots),  \label{eq3.2} \\
   (\partial y_{L}/\partial\check{\mu})_{T} & = & U_{L}L^{\Delta/\nu}(1-j_{2}\check{\rho}+\cdots),  \label{eq3.3}
  \end{eqnarray}
which follow from (\ref{eq2.3}), (\ref{eq1.3}), (\ref{eq1.4}), and (\ref{eq2.12}). On using (\ref{eq2.13}) for $\check{\rho}$ this leads to
  \begin{eqnarray}
   \rhoc\check{\chi}_{N^{3}} & = & e_{1}^{3}U_{L}^{3}L^{(\gamma+\Delta)/\nu}(\partial_{y}^{3}Y) -j_{2}e_{1}^{3}\rhoc^{-1}U_{L}^{4}L^{2\gamma/\nu}  \nonumber \\
   &  & \times\: \left[4(\partial_{y}^{3}Y)(\partial_{y} Y) + 3(\partial_{y}^{2}Y)^{2}\right]  \nonumber \\
  &  &  -\: 3e_{1}^{2}e_{3}U_{L}^{2}D_{L}L^{(\gamma+1)/\nu}(\partial_{x}\partial_{y}^{2}Y) + \cdots, \label{eq3.4}
  \end{eqnarray}
where we recall (\ref{eq2.17}) for $e_{1}$ and $e_{3}$. Using the expansions (\ref{eq2.6}) and then (\ref{eq2.8}), for the scaling functions $Y^{\mbox{\scriptsize\boldmath $\kappa$}}(x_{L},y_{L})$, yields, after some algebra, the defining equation (\ref{eq3.1}) in the form
  \begin{eqnarray}
   &  & \left[ 24 e_{1}Y_{04}^{0} + 24 e_{1}Y_{14}^{0}x_{L} +24e_{1}U_{L4}^{\mbox{\scriptsize c}}Y_{04}^{(4)}L^{-\theta/\nu}\right]y_{L} \nonumber \\
   &  & -\: (3j_{2}+ke_{1})e_{1}\rhoc^{-1}U_{L}L^{-\beta/\nu}\left[2Y_{02}^{0}+2Y_{12}^{0}x_{L}+\cdots\right]^{2}  \nonumber \\
   &  & -\: 3 e_{3}(D_{L}/U_{L})L^{(1-\Delta)/\nu}\left[ 2Y_{12}^{0}+2Y_{22}^{0}x_{L}+\cdots\right]  \nonumber \\
  &  & +\: \cdots = 0.  \label{eq3.5}
  \end{eqnarray}
With the aid of (\ref{eq2.3}) the scaling field $\tilde{h}$ can hence be written in terms of $L$ and $\tilde{t}$ as
  \begin{eqnarray}
   \tilde{h} & = & \mbox{$\frac{1}{24}$}(3j_{2}+ke_{1})/\rhoc Y_{04}^{0}L^{(2-\alpha)/\nu} [2Y_{02}^{0} + 2Y_{12}^{0}D_{L}\tilde{t}L^{1/\nu}  \nonumber \\
  &  &  +\: 2U_{L4}^{\mbox{\scriptsize c}}Y_{02}^{(4)}L^{-\theta/\nu} + \cdots ]^{2} - U_{L4}^{\mbox{\scriptsize c}}Y_{04}^{(4)}\tilde{h}/Y_{04}^{0}L^{\theta/\nu} \nonumber \\
  &  & +\: \cdots.  \label{eq3.6}
  \end{eqnarray}

In order to solve this equation for $\check{\mu}$ as a function of $L$ and $t$, we first write $\check{p}$ in terms of $\check{\mu}$, $t$, and $L$ by using the finite-size scaling equation (\ref{eq2.2}). The expansions (\ref{eq2.8}) for $Y(x_{L},\cdots)$ can then be employed and on solving iteratively for $\check{p}$ we obtain
  \begin{eqnarray}
   \rhoc\check{p} & = & \rhoc(k_{0}t + \check{\mu} + s_{0}L^{-\bar{d}} + \cdots) + Y_{00}^{0}L^{-(2-\alpha)/\nu} \nonumber \\
  &  & +\: D_{L}Y_{10}^{0}[(1-j_{1}k_{0})t - (l_{1}+j_{1})\check{\mu}]L^{-(1-\alpha)/\nu} \nonumber \\
  &  & +\: U_{L4}^{\mbox{\scriptsize c}}Y_{00}^{(4)}L^{-(2-\alpha+\theta)/\nu} + \cdots.  \label{eq3.7}
  \end{eqnarray}
Rewriting (\ref{eq3.6}) yields the reduced chemical potential, $\check{\mu}$, in a similar form from which $\check{p}$ may be eliminated using (\ref{eq3.7}). Solving for $\check{\mu}$ iteratively as a function of $t$ and $L$, finally yields the finite-size $k$-loci in the $(\mu,T)$ plane as
  \begin{eqnarray}
   \check{\mu}^{(k)}(t;L) & = & [ \mu^{(k)}(T;L) - \muc]/k_{\mbox{\scriptsize B}}\Tc \nonumber \\
  & = & \check{\mu}_{1}^{(k)}t + (s_{2}+j_{2}s_{0})L^{-\bar{d}} + M_{1}^{(k)}L^{-(2-\alpha)/\nu} \nonumber \\
  &  & +\: M_{2}^{(k)}L^{-(2-\alpha+\theta)/\nu} + M_{3}^{(k)}tL^{-(1-\alpha)/\nu} \nonumber \\
  &  & +\: \cdots,  \label{eq3.8}
  \end{eqnarray}
where the $M_{j}^{(k)}$ vary linearly with $k$ and are given explicitly in {\bf K}(4.53)-(4.54) while $\check{\mu}_{1}^{(k)}$$\,=\,$$(k_{1}+j_{2}k_{0})/(1-j_{2})$, is actually {\em independent} of $k$ and equal to $\check{\mu}_{\sigma,1}$ which was defined in {\bf I}(3.16) as the (reduced) slope of the phase boundary $\mu_{\sigma}(T)$ {\em at} $T=\Tc$. Notice that, owing to the hyperscaling relation, the $L^{-\bar{d}}$ term here dominates the universal scaling contribution, $L^{-(2-\alpha)/\nu}=L^{-d}$, when $\bar{d}<d$.

Substituting (\ref{eq3.8}) in (\ref{eq3.7}) yields the $k$-loci in the $(p,T)$ plane as
  \begin{eqnarray}
   \check{p}^{(k)}(t;L) & = & [p^{(k)}(T;L)-\pc]/\rhoc k_{\mbox{\scriptsize B}}\Tc  \nonumber \\
  & = & \check{p}^{(k)}_{1}t + [(1+j_{2})s_{0} + s_{2}]L^{-\bar{d}} \nonumber \\
  &  & +\: (M_{1}^{(k)}+Y_{00}^{0})L^{-(2-\alpha)/\nu}  \nonumber \\
  &  & +\: (M_{2}^{(k)} + U_{L4}^{\mbox{\scriptsize c}}Y_{00}^{(4)})L^{-(2-\alpha+\theta)/\nu} \nonumber \\
  &  & +\: (M_{3}^{(k)}+D_{L}Y_{10}^{0}\tau )tL^{-(1-\alpha)/\nu} + \cdots,  \label{eq3.9}
  \end{eqnarray}
where $\check{p}_{1}^{(k)}$$\,=\,$$k_{0}+\check{\mu}_{1}^{(k)}$ is also independent of $k$ and equal to $\check{p}_{\sigma,1}$ [ see {\bf I}(3.12)] while
  \begin{equation}
   \tau = 1-j_{1}k_{0}-(l_{1}+j_{1})(k_{1}+j_{2}k_{0})/(1-j_{2}),  \label{eq3.10}
  \end{equation}
which, in fact, has the same value as $\tau$ in {\bf I}(3.14).

To obtain the $k$-loci in the $(\rho,T)$ plane, we now substitute (\ref{eq3.8}) and (\ref{eq3.9}) into the scaling fields $\tilde{h}$ and $\tilde{t}$ to find
  \begin{eqnarray}
   y_{L} & = & \mbox{$\frac{1}{6}$} (3j_{2}+ke_{1})U_{L}(Y_{02}^{0})^{2}/Y_{04}^{0}L^{\beta/\nu} \nonumber \\
  &  & \times\: [ 1 + 2U_{L4}^{\mbox{\scriptsize c}}Y_{02}^{(4)}/Y_{02}^{0}L^{\theta/\nu} \nonumber \\
  &  & ~~+\: 2D_{L}Y_{12}^{0}\tau tL^{1/\nu}/Y_{02}^{0} + \cdots ],  \label{eq3.11} \\
   x_{L} & = & D_{L}\tau t L^{1/\nu} + \cdots,   \label{eq3.12}
  \end{eqnarray}
and thence can express the generalized densities, $\tilde{\rho}$ and $\tilde{s}$, in (\ref{eq2.11}) in terms of $L$ and $t$. Finally, from (\ref{eq2.13}) we obtain the desired $k$-loci in the $(\rho,T)$ plane as
  \begin{eqnarray}
   \rho^{(k)}(T;L)/\rhoc & = & 1 + B_{1}^{(k)}L^{-2\beta/\nu} + C_{1}^{(k)}L^{-(1-\alpha)/\nu} \nonumber \\
  &  & +\: B_{4}^{(k)}L^{-(2\beta+\theta)/\nu} + \cdots + A_{1}^{(k)}t + \cdots \nonumber \\
  &  & +\: A_{2}^{(k)}L^{-\bar{d}} + B_{5}^{(k)}L^{-(\beta+\theta_{5})/\nu} + \cdots,  \label{eq3.13}
  \end{eqnarray}
where the leading coefficients are
  \begin{eqnarray}
   B_{1}^{(k)} & = & (1-j_{2})(3j_{2}+ke_{1})U_{L}^{2}(Y_{02}^{0})^{3}/3\rhoc^{2}Y_{04}^{0},  \label{eq3.14} \\
   C_{1}^{(k)} & = & -(l_{1}+j_{1})D_{L}Y_{10}^{0}/\rhoc,  \nonumber \\
   B_{4}^{(k)} & = & 3B_{1}^{(k)}U_{L4}^{\mbox{\scriptsize c}}Y_{02}^{(4)}/Y_{02}^{0},  \label{eq3.15} \\
   A_{1}^{(k)} & = & v_{0} + n_{3} + (2q_{0}+n_{0})\check{\mu}_{1}^{(k)} + (n_{0}+2m_{0})\check{p}_{1}^{(k)},  \label{eq3.16}
  \end{eqnarray}
while $A_{2}^{(k)}$ and $B_{5}^{(k)}$, which also entail the {\em nonlinear} scaling-variable coefficients $v_{0}$, $n_{0}$, $m_{0}$, $q_{0}$, $\cdots$ [see {\bf I}(1.4)-(1.6)], are given in {\bf K}(4.62).

Note that the coefficient $A_{1}^{(k)}$ of the leading analytic, $L$-independent term actually coincides with $A_{1}$ in {\bf I}(3.26) which is the amplitude of the linear $t$ term in the coexistence curve diameter. Furthermore, the contribution from the finite-size corrections to the scaling fields, i.e., the $L^{-\bar{d}}$ term in (\ref{eq3.13}) is dominated by $L^{-2\beta/\nu}$, $L^{-(1-\alpha)/\nu}$ and $L^{-(2\beta+\theta)/\nu}$ terms (provided $\bar{d}\geq 2$). When $T=\Tc$, the analytic, $L$-independent part of $\rhoc^{(k)}$ vanishes. The leading correction then decays as $L^{-2\beta/\nu}$ with an amplitude that varies linearly with $k$; this is followed by an $L^{-(1-\alpha)/\nu}$ term whose amplitude does {\em not} depend on $k$. As mentioned in the Introduction, the leading amplitude, $B_{1}^{(k)}$, vanishes, in fact, when $k$ assumes the ``optimal value'' $k_{\mbox{\scriptsize opt}}=-3j_{2}/e_{1} = 3R_{\mu}$, where the Yang-Yang ratio $R_{\mu}$ is defined in Ref.\ [8] and {\bf I} Sec.\ III.E. This value coincides with the one obtained in {\bf I}(4.37) for the thermodynamic limit when it should describe the particular $k$-locus that approaches the critical point ``most directly'' in the $(\rho,T)$ plane.

\subsection{Finite-size {\boldmath $k$}-loci: behavior and applications}
\label{sec3.2}

The near-critical behavior of the finite-size $k$-loci for the hard-core square-well (HCSW) fluid and for the restricted primitive model (RPM) electrolyte is illustrated in Figs.\ 1(a) and (b), respectively.
%%%%%%%%%%%%%%%%%%%%%%%%%%%%%%%%%%
\begin{figure}[ht]
\vspace{-0.2in}
\centerline{\epsfig{figure=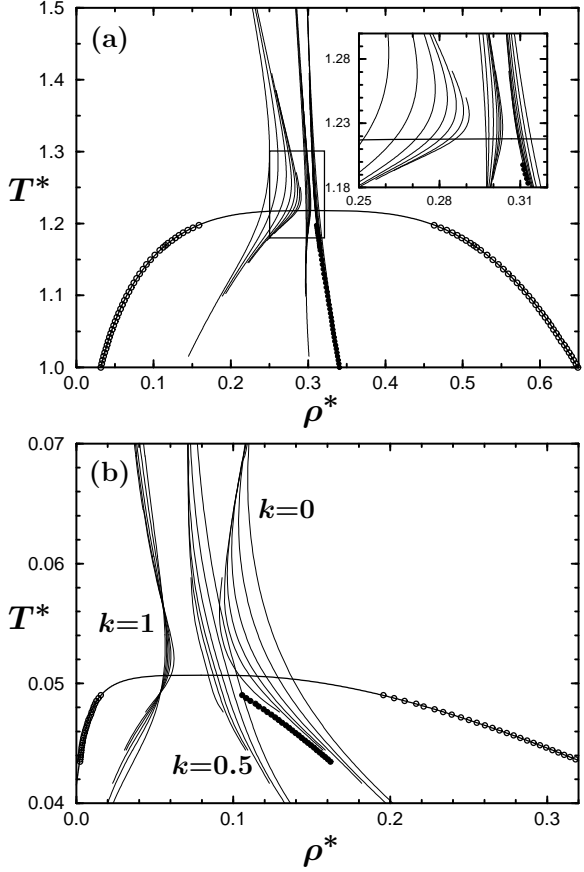,width=4.3in,angle=0}}
\vspace{-0.4in}
\caption{The $k$-loci in the $(\rho,T)$ plane for (a) the hard-core square-well fluid with, from the right, $k=0$, $0.25$ and $1$ where the system sizes $L^{\ast}$ used in the figure are $5$, $6$, $7.5$, $9$, $10.5$, $12$, and $13.5$ (measured in units of the hard-core diameter, $a\equiv\sigma$) [7]; and (b) the restricted primitive model electrolyte with $k=0$, $0.5$ and $1$ where the system sizes shown are $L^{\ast}=6$, $7$, $8$, $9$, $10$, and $12$ [19]. Note that $\rho^{\ast}=\rho a^{3}$ while the reduced temperatures $T^{\ast}$ are defined in Refs [7] and [19] and in Sec.\ V below.}
\end{figure}
%%%%%%%%%%%%%%%%%%%%%%%%%%%%%%%%%%%%%%%%%
 The results shown are based on simulations in periodic cubical boxes of dimensions $L^{\ast}\,$($=\,$$L/a$, where $a$ is the hard-core diameter) upto 13.5 and 12, respectively \cite{ref7,ref19}. The limiting $(L\rightarrow\infty)$ behavior for the same models is shown in Figs.\ 1 and 2 of {\bf I} (while results for a van der Waals fluid are shown in {\bf I} Fig.\ 3). The differences between the HCSW and RPM are quite striking: for the former a value of $k_{\mbox{\scriptsize opt}}$ close to zero or even somewhat negative is suggested while for the RPM one might conclude $k_{\mbox{\scriptsize opt}}\simeq 0.8$. These (inevitably rather uncertain) estimates correspond surprizingly well via $k_{\mbox{\scriptsize opt}}=3R_{\mu}$ with more recent (quite independent) estimates for the Yang-Yang ratio $R_{\mu}$ of $-0.044(3)$ and $+0.26(4)$ for the two models \cite{ref24}.

The result (\ref{eq3.13}) shows that the density estimated {\em at} $T=\Tc$ on the $k$-locus, namely, $\rho^{(k)}(\Tc;L)$, approaches the bulk critical density, $\rhoc$, as, in leading order, $L^{-\psi}$, with $\psi = 2\beta/\nu$ provided the pressure mixing coefficient $j_{2}$ does not vanish. For $(d=3)$ Ising-type criticality this predicts $\psi \simeq 1.03$ whereas for a classical system $\psi = 2$. If $j_{2}$ (and, hence, $R_{\mu}$) vanishes or is numerically small, the next leading term in (\ref{eq3.13}), varying as $L^{-(1-\alpha)/\nu}$, becomes dominant. The exponent $\psi = (1-\alpha)/\nu$ then takes the value $2$ for classical criticality but $\simeq 1.41$ for $(d=3)$ Ising systems.

If a reliable estimate for $\Tc$ is known --- we indicate below [in Sec.IV.C] how this may be found by using the $Q$-loci --- these results can be used in simulations to obtain convincing, {\em unbiased} estimates of the critical density $\rhoc$. By ``unbiased'' we mean that prior knowledge of the critical universality class is {\em not} required. One effective strategy is implemented in Fig.\ 2 for the HCSW fluid
%%%%%%%%%%%%%%%%%%%%%%%%%%%%%%%%%%%%%%
\begin{figure}[ht]
\vspace{-0.2in}
\centerline{\epsfig{figure=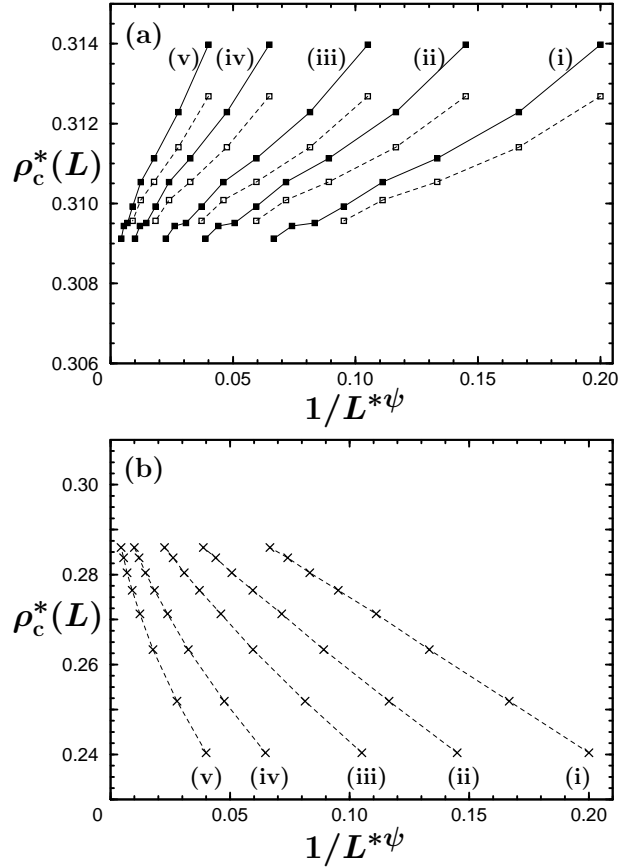,width=4.3in,angle=0}}
\vspace{-0.4in}
\caption{The scaling behavior of $\rhoc^{\ast}(L^{\ast})$ at $T=\Tc$ for (a) the $k$$\,=\,$$0$ locus (solid lines and squares) and (b) the $k$$\,=\,$$1$ locus for a hard-core square-well fluid [7] for trial values of the exponent $\psi$: {\bf (i)} 1.0, {\bf (ii)} 1.2, {\bf (iii)} 1.4, {\bf (iv)} 1.7 and {\bf (v)} 2.0. The dotted lines and open squares in part (a) derive from the $Q$-loci: see Sec.\ IV.A.}
\end{figure}
%%%%%%%%%%%%%%%%%%%%%%%%%%%%%%%%%%%%%
 where $\rhoc^{\ast}(L^{\ast}) \equiv \rho^{(k)}(T$$\,=\,$$\Tc;L^{\ast})a^{3}$ has been plotted for $k=0$ and $k=1$ vs.\ $1/L^{\ast \psi}$ for trial values of the exponent $\psi$ varying from 1 to 2 (which range encompasses both the classical and $(d=3)$ Ising universality classes). For these plots the HCSW estimate $\Tc^{\ast} \equiv k_{\mbox{\scriptsize B}}\Tc/\epsilon \simeq 1.2186$, obtained in Sec.\ IV.C below, has been used. It turns out, however, that $\rho^{(k)}(T;L^{\ast})$ is rather insensitive to $T\simeq \Tc$ so that essentially the same results are obtained if the original Orkoulas {\em et al.} \cite{ref7} estimate (which is about $0.06\%$ lower) is used instead. (Note that this insensitivity is {\em not} realized in the RPM$\,$!)

The straightest plot for $k=1$ [in Fig.\ 2(b)] corresponds to $\psi \simeq 1.0$ which is consistent with Ising behavior (as expected). However, the $k=0$ plots in Fig.\ 2(a) are straightest for $\psi = 1.4\,$-$1.7$: this is also consistent with Ising behavior provided (as seems to be the case) the value of $j_{2}$ is small. Together these plots suggest a critical value of $\rhoc^{\ast}$ in the range $0.3065$ to $0.3080$. To improve the possibilities for extrapolation, the $k=0$ data are combined with data for $k=0.25$ and $0.1$ in Fig.\ 3 and plotted vs.\ $A/(L^{\ast} + l^{\ast})^{\psi}$,
%%%%%%%%%%%%%%%%%%%%%%%%%%%%%%%%%%%%%
\begin{figure}[ht]
\vspace{-0.9in}
\centerline{\epsfig{figure=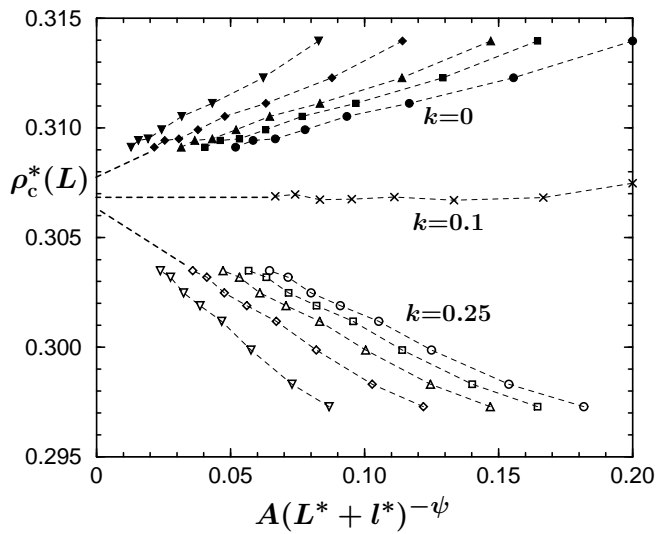,width=3.8in,angle=0}}
\vspace{-1.0in}
\caption{Estimation of the critical density for the HCSW fluid by extrapolation to $L\rightarrow\infty$. The upper solid symbols derive from the $k$$\,=\,$$0$ locus with, from the right, $(\psi,l^{\ast},A)=(1.0,-1.5,0.7)$, $(1.2,-0.5,1.0)$, $(1.4,0,1.4)$, $(1.7,1.0,2.4)$, $(2.0,1.5,3.5)$. The central crosses, from the $k$$\,=\,$$0.1$ locus, have $(\psi,l^{\ast},A)=(1,0,1)$. The lower, open symbols are plotted with, from the right, $(\psi,l^{\ast},A)=(1.0,0.5,1.0)$, $(1.2,2.0,1.7)$, $(1.4,3.0,2.7)$, $(1.7,4.5,5.6)$, and $(2.0,6.0,10.5)$.}
\end{figure}
%%%%%%%%%%%%%%%%%%%%%%%%%%%%%%%%%%%%
 where $A$ is merely a convenient scale factor while the ``shift'' $l^{\ast}$ has been introduced to allow (approximately) for the anticipated higher order corrections. From this figure, we estimate $\rhoc$ for the HCSW fluid (with interaction range $b=1.5a$ \cite{ref7}) as
  \begin{equation}
   \rhoc^{\ast} \equiv \rhoc a^{3} = 0.3068 \pm 0.0007.  \label{eq3.17}
  \end{equation}
This value agrees well with Orkoulas {\em et al.} \cite{ref7} who found $\rhoc^{\ast} = 0.3067 \pm 0.0004$. By the same approach Luijten {\em et al.} \cite{ref19} estimated the critical density of the RPM electrolyte but only to the rather lower precision of $\pm 3\%$ which, however, should be more reliable than other, less systematic and biased methods.

\section{Behavior of the {\boldmath $Q$} parameter and {\boldmath $Q$}-loci}
\label{sec4}

Some time ago Binder \cite{ref20} introduced the dimensionless, finite-system moment ratio, $Q_{L}(T;\langle\rho\rangle_{L}) \equiv \langle m^{2}\rangle^{2}_{L}/\langle m^{4}\rangle_{L}$, defined in a grand canonical ensemble with $m=\rho-\langle\rho\rangle_{L}$, and showed how, in simulations of symmetric systems (where $\rho=\rhoc$ is known), it was particularly useful in locating the critical temperature precisely. Specifically, plots of $Q_{L}(T;\rhoc)$, evaluated on the (symmetric) critical isochore at values of $L$ increased in steps by increments $\Delta L$, display successive intersections at temperatures, say, $T_{Q}^{\Delta L}(L)$, that rapidly approach the limiting, critical temperature $T=\Tc$. At the same time the intersections define a unique and universal critical value \cite{ref21,ref22,ref22a} $Q_{\mbox{\scriptsize c}} = \lim_{L\rightarrow\infty} Q_{L}(\Tc;\rhoc)$. However, the obvious difficulty in attempting to adapt this approach to a {\em non}symmetric fluid system is that the critical density is {\em not} known; nor, in fact, even if $\rhoc$ were known, is it clear that the critical isochore would be the most appropriate locus on which to examine the temperature dependence of $Q_{L}$. Indeed, we will see from our study of $Q_{L}(T;\langle\rho\rangle_{L})$ for general systems that is presented here, that the locus $\rho = \rhoc$, even if known, would not normally be optimal!

To make progress as explained in the Introduction, we define, following \cite{ref19}, the $Q$-loci, $\rho_{Q}(T;L)$, via the isothermal maxima of $Q_{L}(T;\langle\rho\rangle_{L})$ where, it is worth reemphasizing, $\langle\cdot\rangle_{L}$ denotes a grand canonical finite-size average in which $\mu$ is chosen to yield the desired values of the mean density $\langle\rho\rangle_{L}$ (which, of course, is {\em distinct} from what might be considered for a canonical system in which $\rho\equiv N/V$ is directly controlled and does not fluctuate). As seen in Fig.\ 4, for the HCSW fluid and the RPM,
%%%%%%%%%%%%%%%%%%%%%%%%%%%%%%%%%%%%%
\begin{figure}[ht]
\vspace{-0.3in}
\centerline{\epsfig{figure=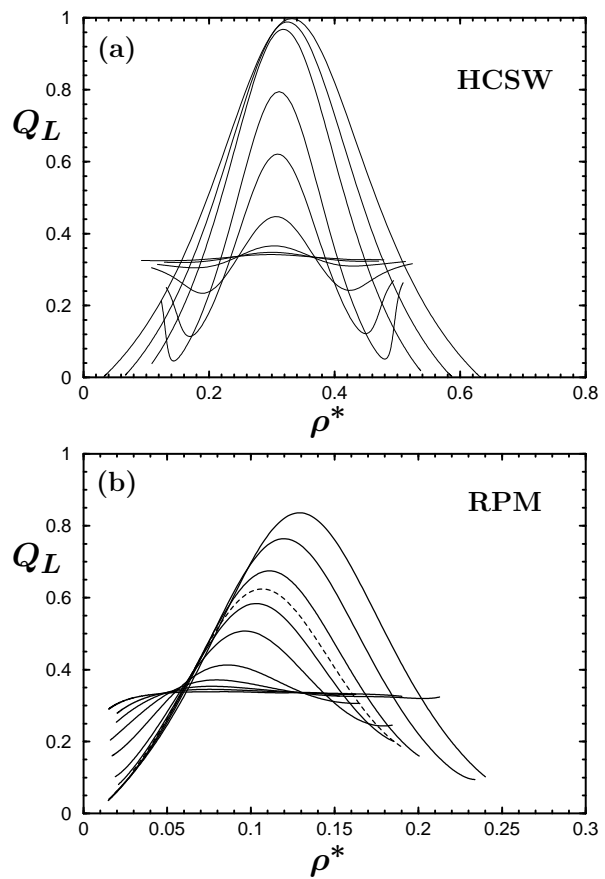,width=4.3in,angle=0}}
\vspace{-0.4in}
\caption{The~moment-ratio~parameter $Q_{L}(T;\rho)$ vs.\ $\rho$ at fixed temperatures (a) for the hard-core square-well fluid at $L^{\ast}$$\,=\,$$10.5$ (from the top, $T^{\ast}$$\,=\,$$1.0$, $1.1$, $1.15$, $1.2$, $1.22$, $1.25$, $1.3$, $1.35,$ and $1.4$): note that $\Tc^{\ast}$$\,\simeq\,$$1.2179$ [7], and (b) for the restricted primitive model electrolyte at $L^{\ast}$$\,=\,$$10$ (from the bottom, the solid-lines are for $1/T^{\ast}$$\,=\,$$13,15$-$19$, $19.5$, $20$, $20.5,$ and $21$); the dashed line is at $\Tc^{\ast}$$\,\simeq\,$$0.050$ [19].}
\end{figure}
%%%%%%%%%%%%%%%%%%%%%%%%%%%%%%%%%%%%
 the ratio $Q_{L}$ at fixed $T$ displays a unique maximum vs.\ density so that $\rho_{Q}(T;L)$ is well defined. In more complex models with, e.g., more than one critical point, the loci will presumably display separate branches or more complex topology; but our concern here is with the behavior of the loci near criticality as $L\rightarrow\infty$, first in the one-phase region above $\Tc$, then through the two-phase region below $\Tc$. In the following section we illustrate the explicit use of these results in simulations.

\subsection{{\boldmath $Q$}-loci above criticality}
\label{sec4.1}

As observed originally by Binder, thermodynamic density fluctuations in a single-phase region of the phase plane should follow a Gaussian distribution when $L\rightarrow\infty$ so that $Q_{L}(T;\langle\rho\rangle_{L})$ for $T>\Tc$ should tend to the constant value $\frac{1}{3}$ as $L$ increases. In practice, as illustrated in Fig.\ 5, the approach at fixed $T$ is nonmonotonic
%%%%%%%%%%%%%%%%%%%%%%%%%%%%%%%%%%%%%
\begin{figure}[ht]
\vspace{-0.9in}
\centerline{\epsfig{figure=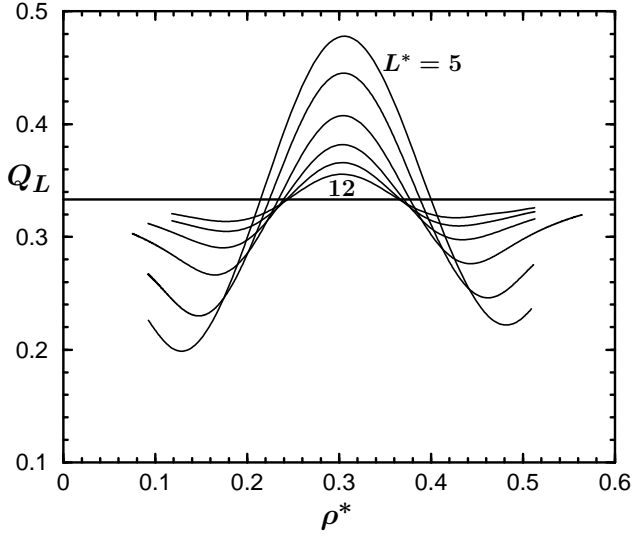,width=3.8in,angle=0}}
\vspace{-1.0in}
\caption{Variation of the moment ratio, $Q_{L}(T;\rho)$, with increasing size for a hard-core square-well fluid at $T^{\ast}$$\,=\,$$1.300$$\,\simeq\,$$1.0674\, \Tc^{\ast}$ [7]; the system dimensions are $L^{\ast}$$\,=\,$$5,6,7.5,9,10.5,$ and $12$. The horizontal solid line represents the single-phase limit $Q_{\infty}$$\,=\,$$\frac{1}{3}$.}
\end{figure}
%%%%%%%%%%%%%%%%%%%%%%%%%%%%%%%%%%%%
 and entails a progression of the $Q$-locus to an apparently well defined limit $\rho_{Q}^{\infty}(T)$.

To estimate the asymptotic behavior of $\rho_{Q}(T;L)$ we may follow the strategy used in studying the $k$-loci. First, in terms of the generalized susceptibilities $\chi_{N^{k}} = (\partial^{k}\bar{p}/\partial\bar{\mu}^{k})_{T}$ with $\bar{p}\equiv p/k_{\mbox{\scriptsize B}}T$ and $\bar{\mu}\equiv \mu/k_{\mbox{\scriptsize B}}T$, note that $Q_{L}$ is equivalent to $V (\chi_{NN})^{2}/\chi_{N^{4}}$. Thence we find
  \begin{equation}
   \left (\frac{\partial Q_{L}}{\partial\bar{\mu}}\right)_{T} = V \frac{\chi_{NN}}{(\chi_{N^{4}})^{2}} [ 2 \chi_{N^{3}}\chi_{N^{4}}-\chi_{NN}\chi_{N^{5}}],   \label{eq4.1}
  \end{equation}
from which, since $\langle\rho\rangle_{L}$ increases monotonically with $\mu$ at fixed $T$, one sees that the $\rho_{Q}$ locus satisfies the equation
  \begin{equation}
   2\check{\chi}_{N^{3}}\check{\chi}_{N^{4}}-\check{\chi}_{NN}\check{\chi}_{N^{5}} = 0.  \label{eq4.2}
  \end{equation}
Here we have employed the reduced susceptibilities $\check{\chi}_{NN}$, $\check{\chi}_{N^{3}}$, etc., introduced in (\ref{eq2.16}) and (\ref{eq3.4}). From (\ref{eq3.4}) we then obtain
  \begin{eqnarray}
   \rhoc\check{\chi}_{N^{4}}& = & e_{1}^{3}U_{L}^{3}L^{(\gamma+\Delta)/\nu}\left[ (\partial_{y}^{4}Y)(\partial y_{L}/\partial\check{\mu})_{T} \right. \nonumber \\
  &  & \left. +\: (\partial_{x}\partial_{y}^{3}Y)(\partial x_{L}/\partial\check{\mu})_{T}\right]  \nonumber \\
   &  & -\: j_{2}e_{1}^{3}\rhoc^{-1}U_{L}^{4}L^{2\gamma/\nu}\left[ 4(\partial_{y}^{4}Y)(\partial_{y}Y) \right.  \nonumber \\
  &  & \left. +\: 10 (\partial_{y}^{3}Y)(\partial_{y}^{2}Y)\right](\partial y_{L}/\partial\check{\mu})_{T} \nonumber \\
  &  & -\: j_{2}e_{1}^{3}\rhoc^{-1}U_{L}^{4}L^{2\gamma/\nu}\left[ 4(\partial_{x}\partial_{y}^{3}Y)(\partial_{y}Y) \right. \nonumber \\
  &  & \left. +\: 4(\partial_{y}^{3}Y)(\partial_{x}\partial_{y}Y)+6(\partial_{x}\partial_{y}^{2}Y)(\partial_{y}^{2}Y)\right](\partial x_{L}/\partial\check{\mu})_{T} \nonumber \\
  &  & -\: 3e_{1}^{2}e_{3}U_{L}^{2}D_{L}L^{(\gamma +1)/\nu}\left[ (\partial_{x}\partial_{y}^{3}Y)(\partial y_{L}/\partial\check{\mu})_{T} \right. \nonumber \\
  &  & \left. +\: (\partial_{x}^{2}\partial_{y}^{2}Y)(\partial x_{L}/\partial\check{\mu})_{T}\right] + \cdots.   \label{eq4.3}
  \end{eqnarray}
Using (\ref{eq3.3}) for $(\partial y_{L}/\partial\check{\mu})_{T}$ and, in that result, (\ref{eq2.13}) for $\check{\rho}$ yields
  \begin{eqnarray}
   \rhoc\check{\chi}_{N^{4}} & = & e_{1}^{4}U_{L}^{4}L^{(\gamma+2\Delta)/\nu}(\partial_{y}^{4}Y) - 5j_{2}e_{1}^{4}\rhoc^{-1}U_{L}^{5}L^{(2\gamma+\Delta)/\nu} \nonumber \\
   &  & \times\: [(\partial_{y}^{4}Y)(\partial_{y} Y) + 2(\partial_{y}^{3}Y)(\partial_{y}^{2}Y)]  \nonumber \\
   &  & -\: 4e_{1}^{3}e_{3}U_{L}^{3}D_{L}L^{(\Delta+\gamma+1)/\nu}(\partial_{x}\partial_{y}^{3}Y) + \cdots.   \label{eq4.4}
  \end{eqnarray}
Similarly, after some algebra, we obtain
  \begin{eqnarray}
   \rhoc\check{\chi}_{N^{5}} & = & e_{1}^{5}U_{L}^{5}L^{(\gamma+3\Delta)/\nu}(\partial_{y}^{5}Y) - j_{2}e_{1}^{5}\rhoc^{-1}U_{L}^{6}L^{2(\gamma+\Delta)/\nu} \nonumber \\
  &  & \times [ 6(\partial_{y}^{5}Y)(\partial_{y}Y) + 15(\partial_{y}^{4}Y)(\partial_{y}^{2}Y) + 10 (\partial_{y}^{3}Y)^{2} ]  \nonumber \\
  &  & -\: 5e_{1}^{4}e_{3}U_{L}^{4}D_{L}L^{(\gamma+2\Delta +1)/\nu}(\partial_{x}\partial_{y}^{4}Y) + \cdots.   \label{eq4.5}
  \end{eqnarray}

Now we may use the leading approximation (\ref{eq2.9}) for the scaling function $Y$ and substitute the expressions (\ref{eq2.16}), (\ref{eq3.4}), (\ref{eq4.4}) and (\ref{eq4.5}) into the $Q$-locus equation (\ref{eq4.2}). This then reduces to
  \begin{equation}
   [ 4(Y_{04}^{0})^{2} - 5Y_{02}^{0}Y_{06}^{0}] y_{L} + 3j_{2}(Y_{02}^{0})^{2}Y_{04}^{0}U_{L}/\rhoc L^{\beta/\nu} + \cdots = 0,   \label{eq4.6}
  \end{equation}
where, for brevity, we have displayed only the leading terms; this, in turn, is readily solved to yield $y_{L}$ on the $Q$-locus as to
  \begin{equation}
    y_{L} \approx -\frac{j_{2}Y_{Q}}{L^{\beta/\nu}}, \hspace{0.1in} Y_{Q} = \frac{3(Y_{02}^{0})^{2}Y_{04}^{0} U_{L}/\rhoc}{4(Y_{04}^{0})^{2}-5Y_{02}^{0}Y_{06}^{0}}.  \label{eq4.6a}
  \end{equation}
To obtain the density, $\check{\rho}$, we appeal to (\ref{eq2.13}) and use (\ref{eq2.11}) for $\tilde{\rho}$ and $\tilde{s}$; then, with (\ref{eq2.6}) and (\ref{eq2.8}) for the scaling functions, and using (\ref{eq4.6a}) for $y_{L}$, we finally obtain the $Q$-locus explicitly as
  \begin{eqnarray}
   \rho_{Q}(T;L)/\rhoc & = & 1 + B_{Q}L^{-2\beta/\nu} + C_{Q}L^{-(1-\alpha)/\nu}  \nonumber \\
   &  & +\: A_{Q}t + \cdots,  \label{eq4.7}
  \end{eqnarray}
where the leading coefficients are
  \begin{eqnarray}
   B_{Q} & = & -\: 2j_{2}(1-j_{2})Y_{Q}Y_{02}^{0}U_{L}/\rhoc, \nonumber \\
   C_{Q} & = & -\: (l_{1} + j_{1})Y_{10}^{0}D_{L}/\rhoc,   \label{eq4.8}
  \end{eqnarray}
while $A_{Q}$ is equal to $A_{1}^{(k)}\equiv A_{1}$, the (reduced) slope of the coexistence curve diameter as given in (\ref{eq3.16}) and {\bf I}(3.26). Note that the leading amplitude, $B_{Q}$, vanishes when $j_{2}=0$. As an explicit example, we present the $Q$-locus for a hard-core square-well fluid \cite{ref7} in Fig.\ 6.
%%%%%%%%%%%%%%%%%%%%%%%%%%%%%%%%%%%%%
\begin{figure}[ht]
\vspace{-0.9in}
\centerline{\epsfig{figure=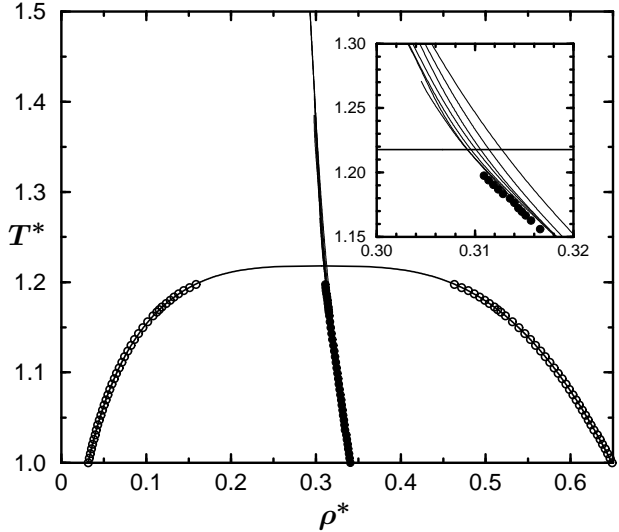,width=3.8in,angle=0}}
\vspace{-1.0in}
\caption{The $Q$-loci in the $(\rho,T)$ plane for a hard-core square-well fluid. From the right, the simulation box dimensions are $L^{\ast}$$\,=\,$$5,6,7.5,9,10.5,12,$ and $13.5$. The estimated critical point shown is $(\rhoc^{\ast},\Tc^{\ast})$$\,=\,$$(0.3067,1.2179)$ [7]; the solid dots represent the estimated coexistence curve diameter.}
\end{figure}
%%%%%%%%%%%%%%%%%%%%%%%%%%%%%%%%%%%%
 Evidently, the loci both above and below $\Tc$ approach the critical point when $L\rightarrow\infty$.

As illustrated by the open squares in Fig.\ 2(a) above, the evaluation of $\rho_{Q}(T;L)$ {\em at} $T=\Tc$ can be used to provide unbiased estimators for the critical density, $\rhoc$, that, in fact, resemble quite closely the sequence provided by the $k$$\,=\,$$0$ loci: see also Fig.\ 5 in Ref.\ \cite{ref19}.

\subsection{Modified or {\boldmath $Q^{(k)}$}-loci}
\label{sec4.2}

It is instructive to define a modified $Q$ parameter, as for $\chi^{(k)}$ in Sec.\ III.A, via
  \begin{equation}
   Q_{L}^{(k)}(T;\langle\rho\rangle_{L}) \equiv Q_{L}(T;\langle\rho\rangle_{L})/\langle\rho\rangle_{L}^{k}.  \label{eq4.10}
  \end{equation}
The modified or $Q^{(k)}$-loci are then defined by the points of isothermal maxima of $Q^{(k)}$ in the $(\rho,T)$ plane. In terms of the reduced susceptibilities, the equation for the locus $\rho_{Q}^{(k)}(T;L)$, becomes
  \begin{equation}
   \check{\rho}[ 2\check{\chi}_{N^{3}}\check{\chi}_{N^{4}}-\check{\chi}_{NN}\check{\chi}_{N^{5}}] = k \check{\chi}_{NN}^{2}\check{\chi}_{N^{4}},  \label{eq4.11}
  \end{equation}
which extends (\ref{eq4.2}). The extension of (\ref{eq4.6}) gains the term $\frac{1}{3}k(1-j_{2})(Y_{02}^{0})^{2}Y_{04}^{0}U_{L}L^{-\beta/\nu}/\rhoc$ on the right hand side. Finally, $\rho_{Q}^{(k)}(T;L)$ is represented by the {\em same} expression (\ref{eq4.7}) [for $k=0$] except that $B_{Q}$ must be replaced by $B_{Q}^{(k)}$ given by
  \begin{equation}
   j_{2}B_{Q}^{(k)} = B_{Q}\left[ j_{2} -\mbox{$\frac{1}{9}$} k(1-j_{2})\right].  \label{eq4.12}
  \end{equation}
This coefficient vanishes when $k=k_{Q} = 9j_{2}/(1-j_{2})=-9R_{\mu}$ which may be contrasted with the ``optimal'' $k$-locus specified by $k_{\mbox{\scriptsize opt}}=3R_{\mu}$ (see Sec.\ III.A). Note that the coefficients $C_{Q}$ and $A_{Q}$ in (\ref{eq4.7}) do {\em not} gain any $k$ dependence although various higher order coefficients will, in fact, depend nonlinearly on $k$.

\subsection{ Behavior of {\boldmath Q} in the two-phase region}
\label{sec4.3}

At fixed $T<\Tc$ the phase transition in the thermodynamic limit is of first-order character with a jump in density from $\rho_{-}(T)$ to $\rho_{+}(T)$ as $\mu$ increases through the phase boundary, $\mu_{\sigma}(T)$. Finite-size scaling theory has been extended to first-order transitions \cite{ref20,ref37,ref38,ref39,ref40} although the main focus previously has been on the dependence as a function of the field $h \propto \mu - \mu_{\sigma}(T)$. Here, motivated by the requirements of simulations, we will enquire more closely into the variation with the density, $\rho$. From this perspective, the crucial feature is that when $\mu\simeq \mu_{\sigma}$ the grand canonical equilibrium distribution function, $P_{L}(\rho;\mu,T)$, exhibits two peaks located at densities near $\rho_{-}(\equiv \rho_{\mbox{\scriptsize vap}})$ and $\rho_{+}(\equiv\rho_{\mbox{\scriptsize liq}})$. For sufficiently large $L$ these peaks can be represented as Gaussians \cite{ref20,ref39,ref40}. Inside the two-phase region one may also need to consider the surface free energy associated with interfaces that separate domains of coexisting phases \cite{ref23,ref37,ref41,ref42}. However, for regularly shaped domains (such as periodic cubes or fixed-shape parallelepipeds) these contributions enter only as exponentially smaller corrections, so they are not considered here. In the case of general fluids the density distribution $P_{L}(\rho)$ has no symmetry: thus for large $L$ in a $d$-dimensional system of fixed regular shape with periodic boundary conditions, we will accept the form \cite{ref23}
  \begin{eqnarray}
  P_{L}(\rho;\mu,T) & \approx & C_{L}\hspace{-0.05in}\left\{ \chi_{-}^{-1/2}\exp [ -\beta (\rho-\rho_{-})^{2}L^{d}/2\chi_{-}] \right.  \nonumber \\
   &  & \left. +\: \chi_{+}^{-1/2}\exp [ -\beta(\rho-\rho_{+})^{2}L^{d}/2\chi_{+}]\right\} \nonumber \\
  &  & \hspace{0.3in} \times \exp[\beta\rho(\mu-\mu_{\sigma})L^{d}],  \label{eq4.13}
  \end{eqnarray}
where $\beta=1/k_{\mbox{\scriptsize B}}T$, while $C_{L}(\mu,T)$ is a normalization constant, and the $\chi_{\pm}(T)$ are the infinite-volume susceptibilities [defined via $\chi = (\partial\rho/\partial\mu)_{T}$] at $\rho=\rho_{\pm}(T)\pm$. This distribution has been set up so that when $\mu=\mu_{\sigma}$ both Gaussians contribute to $P_{L}(\rho)$ with equal weight \cite{ref43}.

To simplify subsequent expressions let us introduce the basic, dimensionless ordering field
  \begin{equation}
    h = [\mu - \mu_{\sigma}(T)]/k_{\mbox{\scriptsize B}}T,  \label{eq4.14}
  \end{equation}
and the average and difference densities and susceptibilities
  \begin{eqnarray}
    \bar{\rho}(T) & = &\mbox{$\frac{1}{2}$}(\rho_{+}+\rho_{-}) \hspace{0.1in} \mbox{and} \hspace{0.1in} \rho_{0}(T)=\mbox{$\frac{1}{2}$}(\rho_{+}-\rho_{-}),  \label{eq4.15} \\
    \bar{\chi}(T) & = & \mbox{$\frac{1}{2}$}(\chi_{+}+\chi_{-}) \hspace{0.1in} \mbox{and} \hspace{0.1in} \chi_{0}(T) = \mbox{$\frac{1}{2}$}(\chi_{+}-\chi_{-}).  \label{eq4.16}
  \end{eqnarray}
Note that $\chi_{0}$ vanishes identically in a symmetric system. For further convenience here we also define the augmented field-dependent densities
  \begin{equation}
   \bar{\rho}^{+}=\bar{\rho}+\bar{\chi}h, \hspace{0.05in} \rho_{0}^{+} = \rho_{0}+\chi_{0}h,\hspace{0.05in} \mbox{and} \hspace{0.05in} \rho_{0}^{(h)} = \rho_{0} + \mbox{$\frac{1}{2}$}\chi_{0}h.  \label{eq4.17}
  \end{equation}

By replacing the summation over discrete density values, $\rho=N/V\geq 0$, by integration over $\rho$ and extending the lower limit to $\rho=-\infty$ (which will entail only an exponentially small error for large $L$), we may compute $\langle\rho\rangle_{L}$ and the moments $\langle m^{n}\rangle_{L}$. This yields
  \begin{equation}
   \langle\rho\rangle_{L}(\mu,T) \approx \bar{\rho}^{+} + \rho_{0}^{+}\tanh (h\rho_{0}^{(h)}L^{d}).  \label{eq4.18}
  \end{equation}
Note that when $h=0$, or $\mu=\mu_{\sigma}(T)$, we have $\langle\rho\rangle_{L}\approx \bar{\rho}(T)$, i.e., the coexistence curve diameter. Likewise we find
  \begin{eqnarray}
   \langle m^{2}\rangle_{L}(\mu,T) & \approx & f_{0} + f_{1}/\beta L^{d},  \label{eq4.19} \\
   \langle m^{4}\rangle_{L}(\mu,T) & \approx & f_{2} + f_{3}/\beta L^{d} + f_{4}/\beta^{2} L^{2d},  \label{eq4.20}
  \end{eqnarray}
where, with
  \begin{equation}
    \Delta\rho \equiv \langle\rho\rangle_{L} - \bar{\rho}^{+} \hspace{0.1in} \mbox{and}\hspace{0.1in} {\cal T} = \tanh (h\rho_{0}^{(h)}L^{d}),  \label{eq4.21}
  \end{equation}
the coefficients may be written
  \begin{eqnarray}
    f_{0} & = & \Delta\rho^{2} + \rho_{0}^{+ 2} - 2\rho_{0}^{+}\Delta\rho {\cal T}, \label{eq4.22} \\
    f_{1} & = & \bar{\chi} + \chi_{0}{\cal T},  \hspace{0.3in} f_{4} = 3(\bar{\chi}^{2}+\chi_{0}^{2}) + 6\bar{\chi}\chi_{0}{\cal T},  \label{eq4.23} \\
    f_{2} & = & \Delta\rho^{4} + 6\rho_{0}^{+ 2}\Delta\rho^{2} + \rho_{0}^{+ 4} \nonumber \\
   &   & -\: 4\rho_{0}^{+}\Delta\rho (\Delta\rho^{2} + \rho_{0}^{+ 2}){\cal T},  \label{eq4.24} \\
    f_{3} & = & 6\bar{\chi}(\Delta\rho^{2} + \rho_{0}^{+ 2}) -12\chi_{0}\rho_{0}^{+}\Delta\rho \nonumber \\
   &   & +\: 6(\chi_{0}\Delta\rho^{2} + 2\bar{\chi}\rho_{0}^{+}\Delta\rho + \chi_{0}\rho_{0}^{+ 2}){\cal T}.  \label{eq4.25}
  \end{eqnarray}   
From these results it is evident that $Q_{L}(\langle\rho\rangle)$ is a ratio of two polynomials of fourth order in $\langle\rho\rangle$ but quadratic in $L^{-d}$.

To examine the two-phase behavior of $Q_{L}$ in the thermodynamic limit, let us define the scaled deviation from the coexistence diameter, $\bar{\rho}(T)$, via
  \begin{equation}
    y \equiv (\rho - \bar{\rho})/\rho_{0},   \label{eq4.26}
  \end{equation}
so that $y \equiv \pm 1$ for $\rho = \rho_{\pm}(T)$. In the first instance we may then, as in \cite{ref19}, set $\mu=\mu_{\sigma}$ (or $h=0$) before allowing $L\rightarrow\infty$. As observed after (\ref{eq4.18}) we then have $\langle\rho\rangle_{L}\rightarrow\langle\rho\rangle_{\infty} = \bar{\rho}$ and ${\cal T} \equiv 0$ in (\ref{eq4.21})-(\ref{eq4.25}). If nonetheless, we identify $\langle\rho\rangle_{L}$ in (\ref{eq4.21}) as $\rho$ in (\ref{eq4.26}) and evaluate $\langle m^{2}\rangle_{\infty}$ and $\langle m^{4}\rangle_{\infty}$ accordingly one is led to
  \begin{equation}
    Q_{\infty}^{\sigma}(T;\rho) = 1 - 4y^{2}/(1+6y^{2}+y^{4}),   \label{eq4.27}
  \end{equation}
which, apart from the superscript $\sigma$ which indicates the limiting procedure adopted, is the result quoted, misleadingly, in \cite{ref19}! Indeed, this can only be the correct limit of $Q_{L}(T;\langle\rho\rangle)$ when $T < \Tc$ if $y=0$, i.e., {\em on} the diameter.

To obtain the true limiting behavior for $-1\leq y \leq 1$, one must first notice that for $\langle\rho\rangle_{L}$ to approach a general value in the interval $(\rho_{-},\rho_{+})$ the thermodynamic limit must be taken with $hL^{d}$ in (\ref{eq4.18}) approaching a finite value that yields $\langle\rho\rangle_{L}\rightarrow\rho$ for the desired value of $y$. This corresponds, in fact, to ${\cal T} \approx \tanh (h\rho_{0}L^{d}) \approx y$ and then yields --- see also \cite{ref23} --- the limiting moments
  \begin{eqnarray}
   \langle m^{2}\rangle_{\infty} & = & \rho_{0}^{2}(1-y^{2}),  \label{eq4.28} \\
   \langle m^{4}\rangle_{\infty} & = & \rho_{0}^{4}(1-y^{2})(1+3y^{2}),  \label{eq4.29}
  \end{eqnarray}
both of which, perhaps surprisingly, {\em vanish} linearly on the phase boundary, i.e., as $y^{2}\rightarrow 1-$. Equally, then \cite{ref23}
  \begin{equation}
   Q_{\infty}(T;\langle\rho\rangle) = (1-y^{2})/(1+3y^{2}) \hspace{0.2in} (T<\Tc),  \label{eq4.30}
  \end{equation}
vanishes linearly on the phase boundary. On the other hand, $Q_{\infty}(T<\Tc)$ takes its maximal value, namely $1$, on the coexistence diameter $(y=0)$. Indeed, the corresponding approach of the $\rho_{Q}(T;\langle\rho\rangle)$ loci below $\Tc$ to the diameter is evident in Fig.\ 6.

To give a graphic impression of the limiting behavior of $Q_{L}(T;\langle\rho\rangle)$ we display in Fig.\ 7, plots constructed using (\ref{eq4.30}) and
%%%%%%%%%%%%%%%%%%%%%%%%%%%%%%%%%%%%%
\begin{figure}[ht]
\vspace{-0.9in}
\centerline{\epsfig{figure=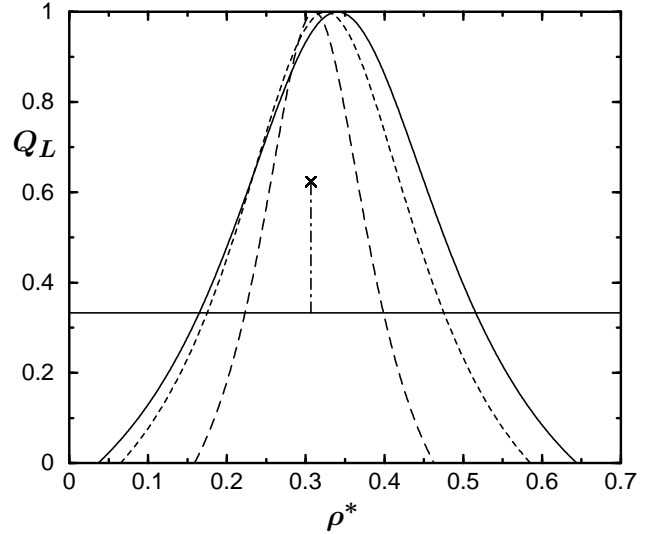,width=3.8in,angle=0}}
\vspace{-1.0in}
\caption{The behavior of the limiting moment ratio, $Q_{\infty}(T;\rho)$, vs.\ $\rho$ at fixed temperatures below $\Tc$ for the hard-core square-well fluid [7]. The solid line, dashed line, and long-dashed line are for $T/\Tc \simeq 0.82,0.90,$ and $0.985$, respectively. The cross is at the critical point $(\Tc^{\ast},\rhoc^{\ast})\simeq (1.218,0.306)$ [7].}
\end{figure}
%%%%%%%%%%%%%%%%%%%%%%%%%%%%%%%%%%%%
 the coexistence curve data for the hard-core square-well fluid [7] at various temperatures below $\Tc$. In addition we have indicated by a cross the anticipated Ising critical point value, $Q_{\mbox{\scriptsize c}}=0.6236 (2)$ \cite{ref22,ref22a}, that we also verify independently below. The horizontal line at $Q=\frac{1}{3}$ describes the limiting single-phase value.

\subsection{Scaling of {\boldmath $Q_{L}(\langle\rho\rangle)$} near coexistence}
\label{sec4.4}

In the one-phase region outside the coexistence curve, i.e., for $y^{2}>1$ [see (\ref{eq4.26})] the result $Q_{\infty}=\frac{1}{3}$ should be recaptured by the analysis based on (\ref{eq4.13}); indeed, the results (\ref{eq4.18})-(\ref{eq4.25}) do confirm this. Thus for $h$ nonzero and $L\rightarrow\infty$, the expression (\ref{eq4.18}) yields
  \begin{equation}
   \langle\rho\rangle_{L} \approx \rho_{\pm} + \chi_{\pm}h - 2(\rho_{0}+\chi_{0}h)e^{-2h\rho_{0}L^{d}},  \label{eq4.31}
  \end{equation}
where the $+$ or $-$ corresponds to $h\gtrless 0$. On substitution in (\ref{eq4.22})-(\ref{eq4.25}) the $L$-independent terms in $\langle m^{2}\rangle_{L}$ and $\langle m^{4}\rangle_{L}$ cancel identically leaving
  \begin{equation}
   \langle m^{2}\rangle_{L} = \chi_{\pm}/\beta L^{d} + O(e^{-2h\rho_{0}L^{d}}),  \label{eq4.32}
  \end{equation}
and, similarly, $\langle m^{4}\rangle_{L} \approx 3\langle m^{2}\rangle_{L}^{2}$, yielding finally
  \begin{equation}
   Q_{L}(T) = \mbox{$\frac{1}{3}$} + O(e^{-2h\rho_{0}L^{d}}),  \label{eq4.33}
  \end{equation}
for $T <\Tc$ and $h$ nonvanishing (but not too large).

Evidently, in the thermodynamic limit, $Q_{\infty}(T;\langle\rho\rangle)$ vanishes as $\rho$ approaches $\rho_{+}$ or $\rho_{-}$ from the two-phase region and then jumps {\em discontinuously} to $\frac{1}{3}$ on entering the single-phase domain. This behavior as $L\rightarrow\infty$ can be seen clearly in grand canonical simulations as illustrated in Fig.\ 8 for the hard-core square-well fluid \cite{ref7}.
%%%%%%%%%%%%%%%%%%%%%%%%%%%%%%%%%%%%%
\begin{figure}[ht]
\vspace{-0.9in}
\centerline{\epsfig{figure=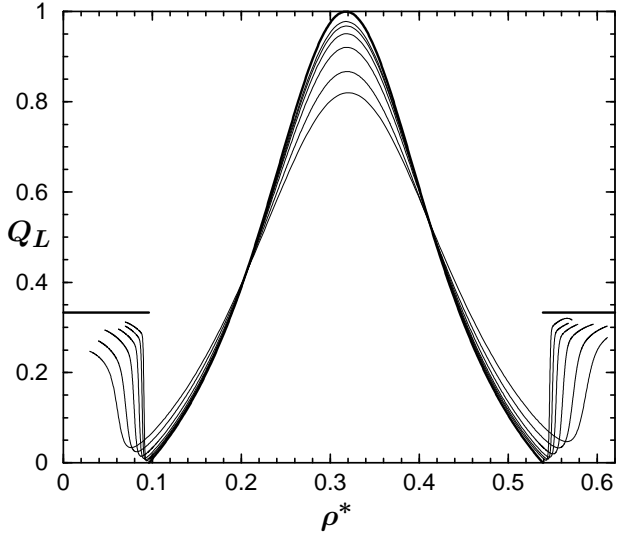,width=3.8in,angle=0}}
\vspace{-1.0in}
\caption{Behavior of $Q_{L}(T;\rho)$ for a hard-core square-well fluid at $T/\Tc \simeq 0.944$ [7]. The thin lines represent simulation data for $L^{\ast} = 5,6,7.5,9,10.5,$ and $12$, while the thick line is the prediction for $L=\infty$ [scaled to the estimated values of $\rho^{\ast}_{+}(T)$ and $\rho_{-}^{\ast}(T)$].}
\end{figure}
%%%%%%%%%%%%%%%%%%%%%%%%%%%%%%%%%%%%
 The predicted limiting behavior is approached rather rapidly at the selected temperature, namely, $\sim$$\, 5\%$ below criticality. However, closer to $\Tc$ and for the RPM the convergence is much slower and less regular as seen in Fig.\ 9 which reports simulations
%%%%%%%%%%%%%%%%%%%%%%%%%%%%%%%%%%%%%
\begin{figure}[ht]
\vspace{-0.3in}
\centerline{\epsfig{figure=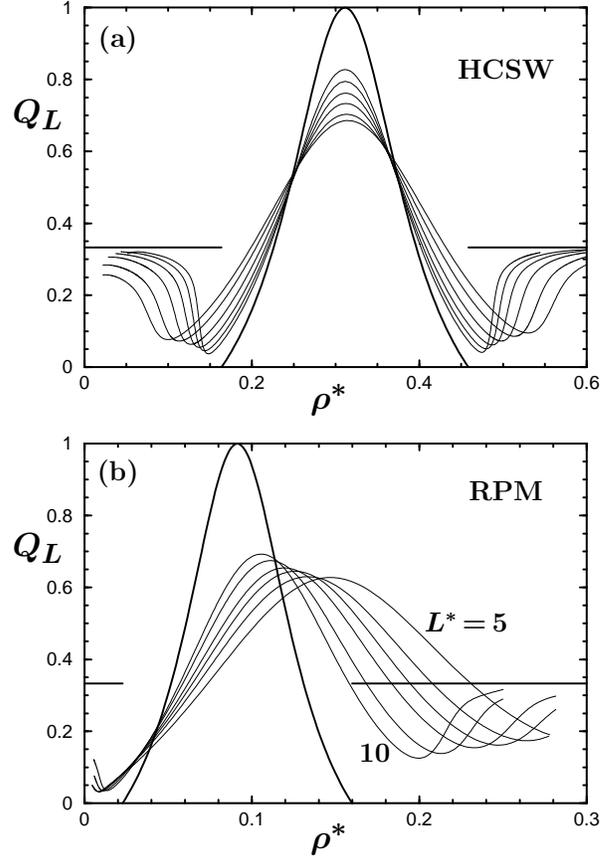,width=4.3in,angle=0}}
\vspace{-0.4in}
\caption{Simulation data for $Q_{L}(T;\rho)$ (a) for the hard-core square-well fluid at $T/\Tc \simeq 0.985$ using the same box sizes $L^{\ast}$ as in Fig.\ 8; (b) for the RPM electrolyte at $T/\Tc \simeq 0.986$ for $L^{\ast}=5$-$10$ [19]. The thick lines represent predictions for the limit $L=\infty$.}
\end{figure}
%%%%%%%%%%%%%%%%%%%%%%%%%%%%%%%%%%%%
 $\sim$$\, 1.5\%$ below the (estimated) critical points. In all cases --- as follows from previous theoretical and simulation-based observations \cite{ref23,ref40,ref41} --- the plots of $Q_{L}(T;\langle\rho\rangle)$ display rounded, but increasingly deep and sharp {\em minima} outside, but approaching, the coexistence curve as $L$ increases. However, the strongly asymmetric and relatively slow approach of the RPM to the limiting behavior is striking. Nevertheless, it turns out that by tracking these minima and suitably extrapolating them on the basis of the present theoretical foundations, remarkably precise estimates of the density jump, $2\rho_{0}(T)=\rho_{+}(T)-\rho_{-}(T)$, and of the diameter, $\bar{\rho}(T)$, can be obtained for both models \cite{ref24}.

In order to understand the minima better let us, for simplicity, consider the symmetric case where $\chi_{+}=\chi_{-}$ so $\chi_{0}\equiv 0$ in (\ref{eq4.13})-(\ref{eq4.16}). After some algebra we obtain from (\ref{eq4.19})-(\ref{eq4.25}) the expression
  \begin{equation}
   Q_{L}(T;\rho) = \frac{ [{\cal X} + (1-{\cal T}^{2})]^{2}}{3{\cal X}^{2} + 6{\cal X}(1-{\cal T}^{2}) + 1 + 2{\cal T}^{2} -3{\cal T}^{4}},  \label{eq4.34}
  \end{equation}
where ${\cal T}(h\rho_{0}L^{d})$ was defined in (\ref{eq4.21}) while
  \begin{equation}
   {\cal X}(T,h;L) = \bar{\chi}(T)/\rho_{0}^{2}(T)k_{\mbox{\scriptsize B}}TL^{d}.  \label{eq4.35}
  \end{equation}
When $L\rightarrow\infty$ so ${\cal X}\rightarrow 0$ and $h\rightarrow 0$ with ${\cal T}^{2}\rightarrow y^{2} <1$, the previous result (\ref{eq4.30}) is recaptured; on the other hand, when $L\rightarrow\infty$ with $h$ fixed and nonzero, one has ${\cal T}^{2}\rightarrow 1$ and (\ref{eq4.33}) is matched. A plot of $Q_{L}$ vs.\ $y$ generated from (\ref{eq4.34}) [see {\bf K} Figs.\ 4.9 and 4.8] quite closely mirrors, except for its precise $y$$\,\Leftrightarrow\,$$-y$ symmetry, the simulations for the HCSW fluid shown in Figs.\ 8 and 9(a): indeed, the HCSW fluid does not deviate drastically from overall symmetry even though it displays some pressure and chemical potential mixing (as discussed above).

For finite $L$, (\ref{eq4.34}) predicts two minima that satisfy
  \begin{eqnarray}
   {\cal T}_{\pm} & = &\pm\; (1+2{\mathcal X})^{1/2}/(1+3{\cal X})^{1/2},  \label{eq4.36} \\
   Q_{\mbox{\scriptsize min}}(T;L) & = & \frac{ {\cal X}(2+3{\cal X}^{2})^{2}}{4+18{\cal X} + 36{\cal X}^{2} + 27 {\cal X}^{3}},  \nonumber \\
   & = & \bar{\chi}/\rho_{0}^{2}k_{\mbox{\scriptsize B}}TL^{d} + O(e^{-2h\rho_{0}L^{d}}).  \label{eq4.37}
  \end{eqnarray}
Thus $Q_{\mbox{\scriptsize min}}(L)$ approaches zero, the limiting value at coexistence, as $L^{-d}$. On the other hand the positions of the minima approach $\rho_{+}$ and $\rho_{-}$ when $L\rightarrow\infty$. In order to find the corresponding $L$ dependence, we first determine $h_{\pm}$ from (\ref{eq4.36}) and (\ref{eq4.21}) obtaining a $(\ln L)/L^{d}$ variation. From (\ref{eq4.18}) we thence find the density minima at
  \begin{eqnarray}
   \rho_{\mbox{\scriptsize min}}^{\pm}(T;L) & = & \rho_{\pm}(T) \pm 2\rho_{0}(T)B_{Q}(T) L^{-d} \nonumber \\
   &  & \times\: [\ln (L^{d}/B_{Q}) - 1 + O(L^{-d}) ],   \label{eq4.38}
  \end{eqnarray}
where the scaling amplitude is
  \begin{equation}
   B_{Q}(T) = k_{\mbox{\scriptsize B}}T\bar{\chi}(T)/4\rho_{0}^{2}(T).  \label{eq4.39}
  \end{equation}
Since this result has been derived only for the symmetric case (although it has wider validity \cite{ref24}) we may replace $\bar{\chi}$ by $\chi_{+}=\chi_{-}$; it is also useful to recall that $2\rho_{0} = \rho_{+}-\rho_{-}=\Delta\rho_{\infty}(T)$ [see (1.11)].

Our discussion of $Q_{L}(T;\langle\rho\rangle)$ below $\Tc$ has, up to this point, been confined to {\em fixed} $T$ and, then, to large enough $L$. On the other hand, when $t=(T-\Tc)/\Tc\rightarrow 0-$ the basic thermodynamic properties entering the expressions for $Q_{L}(T;\langle\rho\rangle)$ and for the minima and their locations will display their standard critical behavior, specifically, $\rho_{0}\sim |t|^{\beta}$, $\bar{\chi}\sim |t|^{-\gamma}$, while $\chi_{0}\sim |t|^{\beta-\gamma}$ [see {\bf I}(3.41,3.42)]. Beyond that, however, the divergence of the correlation length, namely, $\xi \sim a/|t|^{\nu}$, implies that each variable $L$ appearing in the formulas above should, when $t\rightarrow 0-$, be associated with a factor $|t|^{\nu}$. However, the analysis based on the two-Gaussian form (4.13) implicitly assumed that $w \equiv L/\xi\sim L^{\ast}|t|^{\nu}$ was large \cite{ref23,ref37,ref38,ref39,ref40,ref41,ref42}: thus when $t\rightarrow 0-$, we may not simply substitute the expected powers of $t$ in to the expressions so-far derived. On the other hand, the full scaling expression for $Q_{L}$ implied by the basic scaling ansatz (2.2), namely,
  \begin{equation}
   Q_{L}(T;\rho) \approx {\cal Q}(x_{L},Y_{L},y_{L4},y_{L5},\cdots),  \label{eq4.40}
  \end{equation}
must reproduce the expressions obtained here when $w \equiv L/\xi \sim |x_{L}|^{\nu}\rightarrow\infty$ [see (\ref{eq2.3})]. This means that although we cannot hope to derive theoretically an explicit general expression for the scaling function ${\cal Q}(x,y,\cdots)$, or even the scaling forms for the reduced minima, $\rho_{\mbox{\scriptsize min}}^{\pm}(T;L)/\rho_{0}(T)$, we have in essence obtained {\em exact} information about the corresponding scaling behavior! It thus transpire, as shown in \cite{ref24}, that by starting at a temperature below $\Tc$ where $\xi(T)/a = O(1)$, simulation data at increasing $T$ can be used to generate the appropriate scaling functions for $\rho_{\mbox{\scriptsize min}}^{+}(T;L)$ and $\rho_{\mbox{\scriptsize min}}^{-}(T;L)$ and thereby also obtain precise estimates for $\rho_{0}(T)$ and $\bar{\rho}(T)$, i.e., the (limiting) coexistence curve and diameter, even very close to $\Tc$.

\section{ Applications to Simulation}
\label{sec5}

In this section we extend and illustrate the finite-size scaling analysis and the use of the special loci by estimating critical parameters for the hard-core square-well fluid \cite{ref7} and the restricted primitive model electrolyte \cite{ref19} on the basis of grand canonical Monte Carlo simulations. In particular, the $Q$-loci play an important role in determining the critical temperature and indicating the universality class of the models. Once the critical temperature is obtained, we may use the $k$-loci and the $Q$-loci to estimate the critical density $\rhoc$ as already demonstrated in Sec.\ III.B: see Fig.\ 3. To estimate the universal correlation exponent $\nu$ for the RPM, the critical isochore is then utilized.

\subsection{Estimation of {\boldmath $T_{\mbox{\scriptsize\bf c}}$} for the hard-core square-well fluid}
\label{sec5.1}

The HCSW fluid is the simplest continuum model that exhibits realistic gas-liquid separation and criticality. Hard spheres of diameter $a\equiv\sigma$ interact via an attractive square-well pair potential of depth $\varepsilon$ and range $b=\lambda a$. In the simulations discussed here \cite{ref7}, $\lambda$ is taken to be $1.5$ which reasonably represents simple fluids such as argon, etc.\ Reduced temperature and density are defined, as usual, via $T^{\ast}=k_{\mbox{\scriptsize B}}T/\varepsilon$ and $\rho^{\ast}=\rho a^{3}$.

As already observed, for systems with an {\em axis of symmetry}, such as Ising ferromagnets and lattice gases, Binder \cite{ref20,ref39} used the moment parameter ${\cal U}_{L}\equiv (1-1/3Q_{L})$ to estimate critical temperatures (and critical exponents) by evaluating the parameter as a function of $T$ on the axis of symmetry, where, of course, the ordering field, $\tilde{h}$, vanishes identically for {\em all} $L$, and then locating self-intersections. However, asymmetric systems, such as continuum fluids where there is no obvious symmetry axis, pose a crucial question when one aims to apply the same idea: Where should one look? The best choice is, naturally, the locus of ``symmetry'' corresponding to the vanishing of the finite-size ordering field $\tilde{h}(p,T,\mu;L)$. In practice, however, the mixing coefficients $k_{1}$, $j_{2}$, and $s_{2}$ in the ordering field --- see (1.4) and (1.8) --- are not known for such systems so that it is difficult to determine the locus $\tilde{h}=0$ in, say, the $(T,\rho)$ plane.

Furthermore, suppose $Q_{L}$ is calculated along any fixed locus --- such as the critical isochore or even, say, the limiting $Q$-locus $\rho_{Q}^{\infty}(T)$ --- on which $\tilde{h}$ does not vanish but, rather, remains nonzero for any $L$. The contributions to $Q_{L}$ from nonvanishing $\tilde{h}$ may then be gauged by expanding the scaling function ${\cal Q}(x_{L},y_{L},\cdots)$ in (\ref{eq4.40}) about the critical point as
  \begin{eqnarray}
    {\cal Q}(x_{L},y_{L},\cdots) & = & Q_{\mbox{\scriptsize c}}+Q_{1}x_{L}+Q_{2}x_{L}^{2}+Q_{3}y_{L}^{2} \nonumber \\
   &  & +\: Q_{4}y_{L4}+Q_{5}y_{L5}^{2} + \cdots,  \label{eq5.1}
  \end{eqnarray}
where the linear terms $y_{L}$, $y_{L5}$, etc.\ vanish identically in view of the basic symmetry under $y_{L}$$\,\Leftrightarrow\,$$-y_{L}$, $y_{L5}$$\,\Leftrightarrow\,$$-y_{L5}$, etc. Evidently, any small uncertainties in the critical parameters will be enhanced via the scaling combination $y_{L} \propto \tilde{h}L^{\Delta/\nu}$ when $L$ increases. For example, if $\delta\rhoc$ is an error in $\rhoc$, the contribution to $Q_{L}$ will vary as $y_{L}^{2}\sim \delta\rhoc^{2}L^{2\beta/\nu}$ and hence diverge when $L\rightarrow\infty$ thus causing difficulties in extrapolating finite-size data. Explicit calculations reveal the corresponding reduction in precision.

Beyond this issue one finds, by explicit calculations for strongly asymmetric systems like the RPM, that the behavior of $Q_{L}(T)$ {\em on} the critical isochore, $\langle\rho\rangle_{L}=\rhoc$, may not even be monotonic --- as it is on the $\tilde{h}=0$ locus. This adds further uncertainty to interpreting the data.

To overcome these obstacles, we consider the $Q$-loci, $\rho_{Q}(T;L)$, for a fixed $L$ on which it was shown in Sec.\ IV.A that the scaling combination $y_{L}\propto \tilde{h}L^{\Delta/\nu}$ actually {\em decays} as $j_{2}L^{-\beta/\nu}$ when $L$$\,\rightarrow\,$$\infty$: see (4.6). (Thus $\tilde{h}$ vanishes like $j_{2}L^{-(\Delta+\beta)/\nu}$.) Hence, the $Q$-locus can be considered as an ``optimal'' choice for analyzing $Q_{L}$ and estimating $\Tc$. Notice, of course, that in a symmetric system the $Q$-locus reduces to $\tilde{h}=0$ (or, equivalently, to $\rho=\rhoc$). For certain other thermodynamic quantities one might find corresponding optimal loci, such as the $k$-susceptibility-loci, etc. Here we examine $Q_{L}$ evaluated on the $Q$-loci for the HCSW fluid. (For application to the RPM: see \cite{ref19}.)

Generally, one must expect that $Q_{L}$ on a $Q$-locus starts near $Q=\frac{1}{3}$ above $\Tc$ (in the one-phase region); but, since the $Q$-loci in the two-phase region approach the diameter $\bar{\rho}(T)$ [see Fig.\ 6], $Q_{L}$ must then approach unity below $\Tc$ [see (4.30)]. At $T=\Tc$ the $Q$-loci approach the critical point so that $Q_{L}$ on a $Q$-locus must pass through the universal value $Q_{\mbox{\scriptsize c}}$ at some temperature, say $\Tc^{Q}(L)$, that approaches $\Tc$ as $L\rightarrow\infty$. These features are evident in the plots of $Q$ on the $Q$-loci for the HCSW fluid shown in Fig.\ 10.
%%%%%%%%%%%%%%%%%%%%%%%%%%%%%%%%%%%%%
\begin{figure}[ht]
\vspace{-0.9in}
\centerline{\epsfig{figure=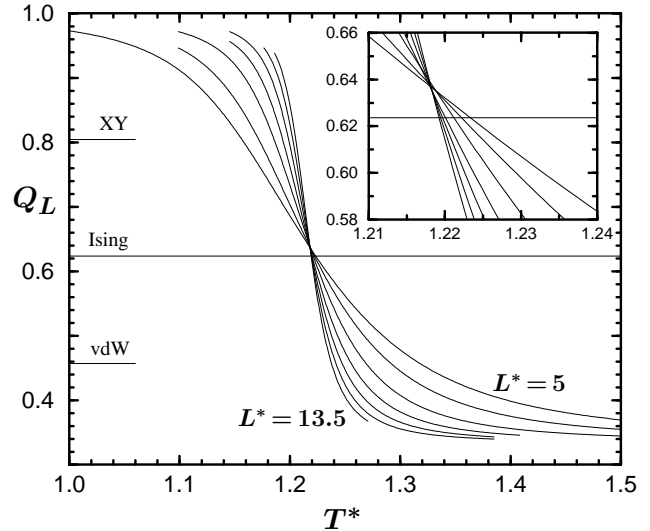,width=3.8in,angle=0}}
\vspace{-1.0in}
\caption{Plots of $Q_{L}(T;\langle\rho\rangle_{L})$ on the $Q$-loci, $\rho_{Q}(T;L)$, for the HCSW fluid providing estimates for $\Tc$ and $Q_{\mbox{\scriptsize c}}$. Classical, XY and Ising values of $Q_{\mbox{\scriptsize c}}$ are marked on the $Q$ axis [21-23]. The system sizes match those in Fig.\ 6.}
\end{figure}
%%%%%%%%%%%%%%%%%%%%%%%%%%%%%%%%%%%%
 Thus all the curves intersect one another near the Ising value $Q_{\mbox{\scriptsize c}}\simeq 0.6236$ (for periodic boundary conditions on a cube \cite{ref21,ref22,ref22a}) strongly confirming that the HCSW fluid belongs to the $(d=3)$-dimensional Ising universality class.

To obtain the asymptotic behavior of $\Tc^{Q}(L)$ for large $L$, we solve the equation
  \begin{equation}
   Q_{L}(T;\rho_{Q}) \approx {\cal Q}(x_{L},y_{L},y_{L4},y_{L5},\cdots)|_{Q} = Q_{\mbox{\scriptsize c}},  \label{eq5.2}
  \end{equation}
where the subscript $Q$ notation denotes evaluation on the $Q$-locus. Substituting the expression (4.7) for $y_{L}$ on the $Q$-loci and using (\ref{eq5.1}), we may solve this equation to obtain
  \begin{eqnarray}
   x_{L} & = & D_{L}\tau t L^{1/\nu} + \cdots,  \nonumber \\
   & \approx & -U_{L4}^{\mbox{\scriptsize c}}Q_{4}/Q_{1}L^{\theta/\nu} - j_{2}^{2}Y_{Q}^{2}Q_{3}/Q_{1}L^{2\beta/\nu},  \label{eq5.3}
  \end{eqnarray}
where $\tau$ was defined in (3.10) and $Y_{Q}$ in (4.7), while the coefficients $Q_{j}$ in (5.1) could also be expressed in terms of the scaling-function expansion coefficients $Y_{lm}^{\mbox{\boldmath\scriptsize $\kappa$}}$. Finally, $\Tc^{Q}(L)$ is given by
  \begin{eqnarray}
   t_{\mbox{\scriptsize c}}^{Q}(L) & \equiv & [\Tc^{Q}(L)-\Tc]/\Tc \nonumber \\
   & = & - P_{1}/L^{(1+\theta)/\nu} - P_{2}/L^{(1+2\beta)/\nu} + \cdots,  \label{eq5.4} \\
   P_{1} & = & Q_{4}U_{L4}^{\mbox{\scriptsize c}}/\tau Q_{1}D_{L}, \hspace{0.1in} P_{2}=j_{2}^{2}Y_{Q}^{2}Q_{3}/Q_{1}\tau D_{L}.  \label{eq5.5}
  \end{eqnarray}
Notice that for $d$$\,=\,$$3$ Ising systems the leading exponents in (\ref{eq5.4}) are $(1+\theta)/\nu\simeq 2.41$ and $(1+2\beta)/\nu\simeq 2.62$, the latter with an amplitude proportional to $j_{2}^{2}$; these large values explain the observed rapid convergence of the $\Tc^{Q}(L)$.

Figure 11 displays $\Tc^{Q}(L)$ versus $L^{-\psi}$ for the HCSW fluid with the predicted Ising value $\psi$$\,=\,$$(1+\theta)/\nu\simeq 2.41$.
%%%%%%%%%%%%%%%%%%%%%%%%%%%%%%%%%%%%%
\begin{figure}[ht]
\vspace{-0.9in}
\centerline{\epsfig{figure=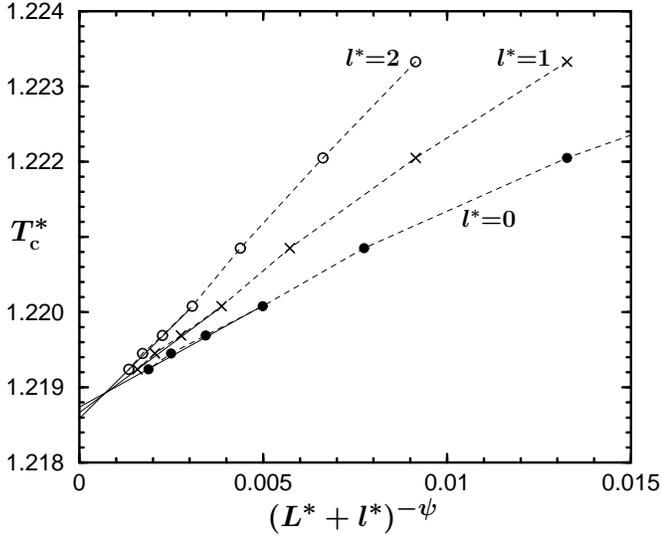,width=3.8in,angle=0}}
\vspace{-1.0in}
\caption{Plots of $\Tc^{Q}(L)$ vs.\ $(L^{\ast} + l^{\ast})^{-\psi}$ with $\psi$$\,=\,$$(1+\theta)/\nu$$\,=\,$$2.41$ to estimate $\Tc^{\ast}$ for the HCSW fluid.}
\end{figure}
%%%%%%%%%%%%%%%%%%%%%%%%%%%%%%%%%%%%
 The small value $R_{\mu}$$\,=\,$$-j_{2}/(1-j_{2})$ [see {\bf I}(3.41)] of about $-0.04$ discussed in Sec.\ III.B indicates that the amplitude $P_{2}$ in (\ref{eq5.4}) is negligible. Thus considering only the leading term is sensible. However, to allow for the various higher order corrections, the small shift parameter $l^{\ast}$ has been introduced. 

From this plot, we estimate the critical temperature for the hard-core square-well fluid to be
  \begin{equation}
   \Tc^{\ast} \simeq 1.2186 \pm 0.0003  \hspace{0.3in} \mbox{(HCSW)}.  \label{eq5.6}
  \end{equation}
This value is about $0.06\%$ higher than the estimate $\Tc^{\ast} \simeq 1.2179\pm 0.0003$ of Orkoulas {\em et al.} \cite{ref7}. For the RPM, Luijten {\em et al.} \cite{ref19} obtained a precision of $\pm 0.04\%$ in estimating $\Tc^{\ast}$ by the same approach.

It is worth stressing that in all these calculations (and those described above and below) it has been imperative to use extensive histogram reweighting procedures \cite{ref44} in order to precisely determine intersections of loci, maxima and minima, etc. It is clear that without sufficient precision and, indeed, accuracy in calculating finite-size properties, extrapolation procedures are doomed to failure or, worse, seriously misleading estimates.

\subsection{Estimation of {\boldmath $Q_{\mbox{\scriptsize\bf c}}$}}
\label{sec5.2}

There seems little serious doubt on the basis of Fig.\ 10 (as well, of course, as on previous evidence \cite{ref7}) that criticality in the HCSW fluid is of short-range Ising type. In other cases, however, one may well desire to estimate $Q_{\mbox{\scriptsize c}}$, and hence resolve the universality class, in unbiased fashion. In that situation the successive intersections of plots of $Q_{L}$ on the $Q$-loci for increasing sequences of $L$ values may be useful. Accordingly, let us define $T_{Q}^{\Delta L}(L)$ and $Q_{Q}^{\Delta L}(L)$ as the intersections of a plot of $Q_{L}(T)$ on the $\rho_{Q}(L;T)$ locus with a plot of $Q_{L-\Delta L}(T)$ on the $\rho_{Q}(L-\Delta L;T)$ locus and ask for the asymptotic behavior as $L$ increases at fixed, small $\Delta L$.

The analysis follows the lines of the previous section except that (\ref{eq5.2}) is replaced by
  \begin{equation}
   {\cal Q}(x_{L},y_{L},\cdots)|_{Q} - {\cal Q}(x_{L-\Delta L},y_{L-\Delta L},\cdots)|_{Q} \approx 0.  \label{eq5.7}
  \end{equation}
For the temperature intersections we find
  \begin{eqnarray}
   [T_{Q}^{\Delta L}(L)-\Tc]/\Tc & = & \theta P_{1}/L^{(1+\theta)/\nu} \nonumber \\
  &  & +\: 2\beta P_{2}/L^{(1+2\beta)/\nu} + \cdots,  \label{eq5.8}
  \end{eqnarray}
which, in leading order, is {\em independent} of $\Delta L$. The coefficients $P_{1}$ and $P_{2}$ are the same as those defined in (\ref{eq5.5}), that enter (\ref{eq5.4}), namely the asymptotic result for $t_{\mbox{\scriptsize c}}^{Q}(L)$, the intersections with $Q_{\mbox{\scriptsize c}}$. However, the approach takes place from the opposite side, and since $\theta \simeq 0.52$ and $2\beta\simeq 0.65$, the amplitudes are smaller. For these reasons one might well prefer to use the successive intersections: however, a little reflection shows that they place greater demands on the precision and reliability of the simulations.

Unfortunately, the convergence of the estimates for $Q_{\mbox{\scriptsize c}}$ is not as rapid. We find
  \begin{eqnarray}
   Q_{Q}^{\Delta L}(L) & \approx & Q_{\mbox{\scriptsize c}} + (1+\theta)Q_{4}U_{L4}^{\mbox{\scriptsize c}}/L^{\theta/\nu} \nonumber \\
  &  & +\: (1+2\beta)j_{2}^{2}Q_{3}Y_{Q}^{2}/L^{2\beta/\nu},  \label{eq5.9}
  \end{eqnarray}
where for Ising-type systems the exponents are $\theta/\nu \simeq 0.83$ and $2\beta/\nu \simeq 1.04$. This slower convergence may be the reason why the successive intersections seen in the inset in Fig.\ 10 suggest a limit some $1$ or $2\%$ higher than the established Ising value \cite{ref21,ref22,ref22a}. However, since no special efforts were originally made \cite{ref7} to gather HCSW data optimal for evaluating $Q$ and the $Q$-loci, one must also suspect the possibility of inadequate simulation accuracy. By contrast, the central unbiased estimate for $Q_{\mbox{\scriptsize c}}$ for the RPM (on which considerable effort was focussed) captured the Ising value precisely within uncertainties of only $\pm 0.3\%$ \cite{ref19}.

\subsection{Estimating the correlation exponent}
\label{sec5.3}

Of basic importance and value in determining the universality class of a model is the correlation length exponent $\nu$. As already frequently stressed, this enters in finite-size systems via the combination $L|\tilde{t}|^{\nu}$ which opens many routes to the estimation of $\nu$. For example, the scaling of $Q_{L}$ on the $Q$-locus should satisfy
  \begin{equation}
   Q_{L}\mbox{\large\bf $($}T;\rho_{Q}(T;L)\mbox{\large\bf $)$} - Q_{\mbox{\scriptsize c}} \approx \Delta {\cal Q}(tL^{1/\nu}). \label{eq5.10}
  \end{equation}
From this it follows that the {\em derivatives}, $\partial Q_{L}(T;\rho_{Q}(T;L))/\partial T$, evaluated at $\Tc$ {\em or} at $\Tc^{Q}(L)$ {\em or} at $T_{Q}^{\Delta L}(L)$, etc., will all, in leading order, diverge as $L^{1/\nu}$. However, obtaining these derivatives accurately is a difficult computational task. Furthermore, the corrections to the leading behavior are likely to be quite significant (owing, in particular, to the strongly nonlinear variation of $\Delta {\cal Q}(x)$ which must saturate at constant values of order unity when $x\rightarrow\pm\infty$).

To provide a robust method of estimating $\nu$ from simulations {\em above criticality} --- which are intrinsically easier to bring to equilibrium than simulations closer to or below $\Tc$ --- Orkoulas {\em et al.} \cite{ref7} introduced various ``estimator functions,'' ${\cal Y}_{j}(T,\mu)$. When evaluated in the thermodynamic limit on a critical locus, say $\zeta$, that approached the critical point from above, these diverged as $t\rightarrow 0$; but in a finite system they exhibited rounded maxima {\em above} $\Tc$ at temperatures $T_{j}(L)$. For suitable loci, $\zeta$, the $T_{j}(L)$ must approach $\Tc$ as $L^{-1/\nu}$. Then Orkoulas {\em et al.} considered {\em unbiased} exponent estimators, {\em independent} of the unknown (or known) value of $\Tc$. Specifically, for a pair ${\cal Y}_{j}$ and ${\cal Y}_{k}$, they measured $\Delta T_{jk}=T_{j}(L)-T_{k}(L)$ and computed sequences
  \begin{equation}
   \Lambda_{jk} \equiv \left[ 1 - \frac{\Delta T_{jk}(L+\Delta L)}{\Delta T_{jk}(L)}\right]\frac{L}{\Delta L} \rightarrow \frac{1}{\nu},  \label{eq5.11}
  \end{equation}
as $L\rightarrow\infty$. By using estimates for the critical isochore, Orkoulas {\em et al.} \cite{ref7} estimated $\nu$ for the HCSW fluid and confirmed its Ising-type character. They also checked that, within the available precision, the results for $\nu$ were not sensitive to the estimate for $\rhoc$.

However, this method is relatively demanding in that the differences, $T_{j}(L)-T_{k}(L)$, must be obtained to relatively high precision. For the RPM --- which is much harder to simulate reliably than the HCSW fluid even above $\Tc$ --- this proved a stumbling block. In addition, while relative insensitivity to the estimate of $\rhoc$ could reasonably be expected, the very strong asymmetry and the likelihood of strong pressure mixing (since confirmed \cite{ref24}) made the choice of critical locus more questionable. Would the critical isochore still be satisfactory?

At issue in this latter question is that, as a result of pressure mixing, the estimator functions ${\cal Y}_{j}(T)$ pick up contributions varying with the fields $\tilde{h}$ and $\tilde{p}$ on the locus $\zeta$, say $\rho=\rhoc$. This question is partly resolved by the analysis of {\bf I} Sec.\ IV.D which shows that on the critical isochore [and, by extension, on any locus behaving asymptotically as $(\rho-\rhoc)$$\,\approx\,$$ct$$\,\rightarrow\,$$0$] one has $\tilde{h}$$\,\sim\,$$|t|^{1-\alpha+\gamma}$ and $\tilde{p}$$\,\sim\,$$|t|^{2-\alpha}$. The associated correction exponents are sufficiently large $(\gtrsim 2)$ that they are of little practical concern relative to the unavoidable leading correction-to-scaling terms varying as $t^{\theta}$. In a finite system a discussion along the lines leading to (\ref{eq5.9}) (that invokes the analog of (4.7) for $y_{L}$ on the isochore) is appropriate; but, as in (\ref{eq5.9}), the extra terms to be anticipated, varying as $L^{-2\beta/\nu}$, are of higher order than the leading $L^{-\theta/\nu}$ corrections. Nevertheless, it may be of value, as suggested in \cite{ref9}, to use as the locus $\zeta$ a ``theta locus'' defined via 
  \begin{equation}
   \rho_{\vartheta}(T) = \rhoc [ \vartheta + (1-\vartheta)(\Tc/T)],  \label{eq5.12}
  \end{equation}
where a most favorable value of $\vartheta$ might be one chosen to approximate an optimal $k$-locus or $Q^{(k)}$-locus.

For the RPM a second problem arises which we explain here and then deal with explicitly. For completeness we recall that the restricted primitive model electrolyte consists of $N$$\,=\,$$2N_{+}$ hard spheres of diameter $a$$\,\equiv\,$$\sigma$, of which $N_{+}$ carry a charge $+q_{0}$ and $N_{-}(=N_{+})$ a charge $-q_{0}$. The pairwise Coulomb potential is $\pm q_{0}^{2}/Dr$ for two like/unlike charges at separation $r$. Appropriate reduced variables are
  \begin{equation}
   T^{\ast} = k_{\mbox{\scriptsize B}}TDa/q_{0}^{2}, \hspace{0.3in} \rho^{\ast} = \rho a^{3}.  \label{eq5.13}
  \end{equation}

Orkoulas {\em et al.} introduced {\em twelve} estimator functions ${\cal Y}_{j}$ $(j=1,\cdots,12)$ \cite{ref7}. The simplest, ${\cal Y}_{1}=C_{V}$, was the constant volume heat capacity. But for the RPM this displays maxima fairly far {\em below} $\Tc$ which, moreover, are not easy to locate precisely \cite{ref45,ref46,ref47}. With $\Theta = 1/T^{\ast}$ Orkoulas {\em et al.} defined ${\cal Y}_{2}= (\partial C_{V}/\partial\Theta)_{\rho}$: this function has a local extremum, $T_{2}^{+}(L)$, above $\Tc$ which varies fairly regularly as $L$ increases: see Fig.\ 12.
%%%%%%%%%%%%%%%%%%%%%%%%%%%%%%%%%%%%%
\begin{figure}[ht]
\vspace{-0.9in}
\centerline{\epsfig{figure=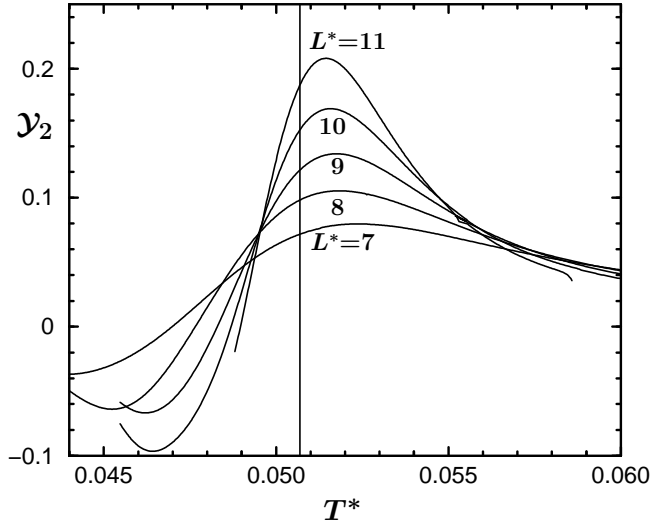,width=3.8in,angle=0}}
\vspace{-1.0in}
\caption{The estimator function ${\cal Y}_{2}(T)$$\,=\,$$(\partial C_{V}/\partial \Theta)$ with $\Theta$$\,=\,$$1/T^{\ast}$ on the critical isochore $(\rhoc^{\ast} \simeq 0.079)$ of the RPM electrolyte (at a $\zeta$$\,=\,$$5$ discretization level [19]). The vertical line marks the estimated critical point at $\Tc^{\ast} \simeq 0.05069$ [19].}
\end{figure}
%%%%%%%%%%%%%%%%%%%%%%%%%%%%%%%%%%%%
 On the other hand, in the case of the RPM the functions ${\cal Y}_{3}, \cdots, {\cal Y}_{6}$ prove to have maxima close to but {\em below} $\Tc$. The function ${\cal Y}_{7}$, a modified susceptibility, displays {\em no maxima} on the critical isochore in the range $0.045 \leq T^{\ast} \leq 0.070$. The remaining functions ${\cal Y}_{8}$ to ${\cal Y}_{12}$ do display extrema above $\Tc$ but their behavior is not very smooth for the accessible values of $L^{\ast}$!

Accordingly, new estimator functions were sought. After some investigation two further acceptable functions were found, namely,
  \begin{equation}
   {\cal Y}_{4}^{\prime} \equiv \left(\frac{\partial^{2}\langle m^{2}\rangle^{1/2}}{\partial\Theta^{2}}\right)_{\rho}, \hspace{0.1in} {\cal Y}_{6}^{\prime} \equiv \left(\frac{\partial^{2}\langle m^{6}\rangle^{1/6}}{\partial\Theta^{2}}\right)_{\rho},  \label{eq5.14}
  \end{equation}
where $m$$\,=\,$$(N$$\,-\,$$\langle N\rangle)/V$. The behavior of these functions resembles that shown for ${\cal Y}_{2}(T)$ in Fig.\ 12 although for the same values of $L$ the maxima lie further from $\Tc$: see {\bf K} Figs.\ 4.15 and 4.16.

Finally, we must accept that neither the quantity nor the quality of the  obtainable RPM data suffice to implement the recipe (\ref{eq5.11}). Instead, we accept the biased estimators
  \begin{equation}
   \Lambda_{j} = \left[ 1-\frac{T_{j}(L+\Delta L)-\Tc}{\Delta T_{j}(L)-\Tc}\right]\frac{L^{\ast} + l^{\ast}}{\Delta L^{\ast}} \rightarrow\frac{1}{\nu},  \label{eq5.15}
  \end{equation}
which require a value for $\Tc$: that we take from the study of $Q$ on the $Q$-loci as in Fig.\ 10 \cite{ref19}. The shift parameter $l^{\ast}$ allows, as in Fig.\ 11, for higher order terms in the behavior of the $T_{j}(L)$. Extrapolation vs.\ $1/L$, as illustrated in Fig.\ 13,
%%%%%%%%%%%%%%%%%%%%%%%%%%%%%%%%%%%%%
\begin{figure}[ht]
\vspace{-0.9in}
\centerline{\epsfig{figure=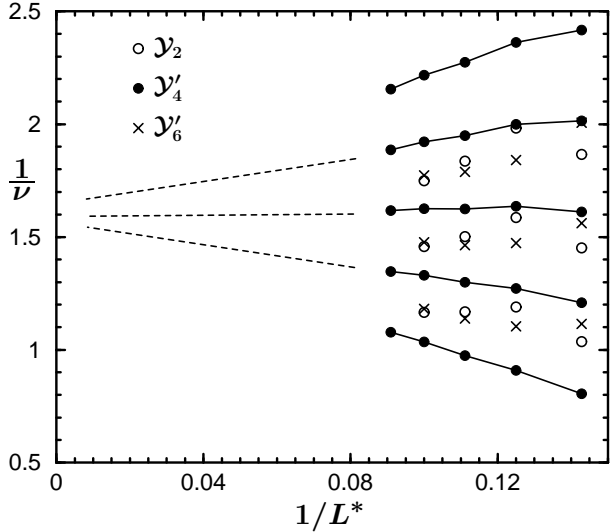,width=3.8in,angle=0}}
\vspace{-1.0in}
\caption{Plots of the estimators (5.14) for the exponent $1/\nu$ for the RPM using ${\cal Y}_{2}$ (open circles) and ${\cal Y}_{6}^{\prime}$ (crosses) with $l^{\ast}$$\,=\,$$2,0,$ and $-2$ from the top downwards, and ${\cal Y}_{4}^{\prime}$ (solid circles) with $l^{\ast}$$\,=\,$$5,3,1,-1,$ and $-3$: see text and (5.14).}
\end{figure}
%%%%%%%%%%%%%%%%%%%%%%%%%%%%%%%%%%%%
 yields
  \begin{equation}
   \nu = 0.63 \pm 0.03  \hspace{0.3in} \mbox{(RPM)}.  \label{eq5.16}
  \end{equation}
This value (previously reported but not justified \cite{ref19}) supports the conclusion that, despite the infinite range of the ionic forces underlying the model, it behaves, as regards phase separation and criticality, like a short-range Ising-type system.

\section{Full Scaling in the Canonical Ensemble}
\label{sec6}

In the thermodynamic limit for regular systems there is a full equivalence between the different ensembles. Consequently a ``canonical description'' in terms of the Helmholtz free energy density $f(\rho,T) = \lim_{L\rightarrow\infty} F_{N}(V,T)/V$ with $\rho = \lim_{L\rightarrow\infty} (N/V)$, is as valid and provides the same information as the grand canonical viewpoint based on $p(T,\mu)$ that we have so far adopted. Similarly, as observed in the Introduction, in leading order the canonical scaling form (1.13), which invokes the scaled combination $z \propto m/|t|^{\beta}$, is equivalent to the grand canonical form (2.4) which entails $y \propto h/|t|^{\Delta} \propto [\mu - \mu_{\sigma}(T)]/|t|^{\Delta}$. However, in higher orders the necessity for field mixing via (1.1)-(1.4) complicates matters. Specifically, whereas the full scaling fields $\tilde{t}$, $\tilde{\mu}$ and $\tilde{p}$ are generally {\em nonsingular} functions of the underlying scaling fields, $t$, $\mu$, and $p$ (unless renormalization group ``resonances'' arise \cite{ref27}), this is no longer the case for the canonical variables $\tilde{m}$, $\tilde{t}$ and $\tilde{f}$. Here we derive some of these complications that arise canonically, first in the thermodynamic limit in the presence of pressure mixing, then in finite systems. In the latter case we wish, in particular, to understand the asymptotics of the finite-size, classical-type critical points that may be identified in canonical simulations: see, e.g., \cite{ref7,ref19}.

\subsection{ Thermodynamic limit}
\label{sec6.1}

By standard thermodynamics for infinite systems the Helmholtz free energy density is given by
  \begin{equation}
   f(\rho,T) = \rho\mu - p,  \label{eq6.1}
  \end{equation}
where $\mu$ and $p$ are understood to be re-expressed in terms of the density via $\rho = (\partial p/\partial\mu)_{T}$. It is straightforward to introduce the reduced variables $\check{\rho}$, $\check{p}$ and 
  \begin{equation}
   \check{\mu} = e_{0}(\tilde{h} + j_{2}\tilde{p} + e_{4}t +\cdots),   \label{eq6.2}
  \end{equation}
via (1.1) and (2.12), and convenient to recall (2.17), for $e_{1}$ and $e_{3}$, and, further, to write
  \begin{equation}
   e_{0} = e_{1}^{-1},\hspace{0.1in} e_{2}=j_{1}+j_{2}l_{1}, \hspace{0.1in} e_{4}=k_{1}+j_{2}k_{0}.  \label{eq6.3}
  \end{equation}

Now we must address the choice of general canonical scaling variables. We wish, first, to allow for the leading correction-to-scaling terms which are expressed in terms of $\tilde{t}$ both for infinite and finite systems in (2.2)-(2.4). Accordingly, it seems appropriate to adopt $\tilde{t}$ also canonically, although it will need to be re-expressed in terms of $m$ in place of $\check{\mu}$.

Similarly, it seems clear that the general scaling field $\tilde{m}$ should be chosen conjugate to the general ordering field $\tilde{h}$. Thus we adopt
  \begin{equation}
   \tilde{m} = (\partial\tilde{p}/\partial\tilde{h})_{\tilde{t}},  \label{eq6.4}
  \end{equation}
which is identical to the scaling density $\tilde{\rho}$ that was introduced in (2.10) along with the scaling entropy $\tilde{s}$.

With these variables in hand we can rewrite (\ref{eq6.1}) as
  \begin{equation}
   \check{f}(\rho,T) \equiv f(\rho,T)/\rhoc k_{\mbox{\scriptsize B}}\Tc = \check{f}_{0}(\rho,T) + \tilde{f}(\rho,T),  \label{eq6.5}
  \end{equation}
where the nonsingular background term may be expanded as
  \begin{equation}
   \check{f}_{0}(\rho,T) = \check{f}_{\mbox{\scriptsize c}} + \bar{\mu}_{\mbox{\scriptsize c}}\check{\rho} - k_{0}t + e_{0}e_{4}\check{\rho}t + \cdots,  \label{eq6.6}
  \end{equation}
with $\check{f}_{\mbox{\scriptsize c}}=(\rhoc\mu_{\mbox{\scriptsize c}}-p_{\mbox{\scriptsize c}})/\rhoc k_{\mbox{\scriptsize B}}\Tc$ and $\bar{\mu}_{\mbox{\scriptsize c}}=\muc/k_{\mbox{\scriptsize B}}\Tc$. On the other hand, the singular contribution becomes
  \begin{equation}
   \tilde{f}(\rho,T) = \tilde{m}\tilde{h} - \tilde{p} + j_{2}\tilde{m}\tilde{p} - e_{0}e_{3}\tilde{s}\tilde{h} - j_{2}e_{0}e_{3} \tilde{s}\tilde{p} + \cdots,  \label{eq6.7}
  \end{equation}
in which the presence of the coefficient $j_{2}$ makes clear how pressure mixing enters.

Our aim now is to express $\tilde{f}$ in terms of the general canonical scaling combinations
  \begin{equation}
   z = \tilde{m}/\tilde{B}|\tilde{t}|^{\beta}, \hspace{0.1in} y_{4} = U_{4}|\tilde{t}|^{\theta}, \hspace{0.1in} y_{5}=U_{5}|\tilde{t}|^{\theta_{5}}, \hspace{0.1in} \cdots,  \label{eq6.8}
  \end{equation}
where $\tilde{B} = QU$: see (2.4), (2.5) and accompanying text. To that end, from (2.4) and (\ref{eq6.4}) we first obtain
  \begin{equation}
   z = W_{\pm}^{\prime}(y;y_{4},y_{5},\cdots) \hspace{0.1in} \mbox{with} \hspace{0.1in} y = U\tilde{h}/|\tilde{t}|^{\Delta},  \label{eq6.9}
  \end{equation}
for $\tilde{t}\gtrless 0$, where $W_{\pm}^{\prime}(y;\cdots) = \partial W_{\pm}/\partial y$. Inverting this expression yields
  \begin{equation}
   \tilde{h} = U^{-1}|\tilde{t}|^{\Delta}F_{\pm}^{\mu}(z;y_{4},y_{5},\cdots),  \label{eq6.10}
  \end{equation}
where the scaling functions $F_{\pm}^{\mu}(z)$ are the inverses of the $W_{\pm}^{\prime}(y)$. From (2.4), we hence find
  \begin{equation}
   \tilde{p}= Q|\tilde{t}|^{2-\alpha}F_{\pm}^{p}(z;y_{4},y_{5},\cdots),  \label{eq6.11}
  \end{equation}
in which the new scaling functions are defined by
  \begin{equation}
   F_{\pm}^{p}(z;y_{4},y_{5},\cdots) = W_{\pm}\mbox{\boldmath\large $($}F_{\pm}^{\mu}(z;y_{4},\cdots);y_{4},\cdots\mbox{\boldmath\large $)$}.  \label{eq6.12}
  \end{equation}

Then, rearranging (1.2)-(1.4) and substituting yields the {\em canonical thermal scaling field} as
  \begin{eqnarray}
   \tilde{t} & = & \tau t - e_{0}e_{3}\tilde{h} - e_{0}e_{2}\tilde{p}+ \cdots, \nonumber \\
  & = & \tau t - (e_{0}e_{3}/U)|\tilde{t}|^{\Delta}F_{\pm}^{\mu}(z;y_{4},\cdots) \nonumber \\
  &   & -\: e_{0}e_{3}Q|\tilde{t}|^{2-\alpha}F_{\pm}^{p}(z;y_{4},\cdots) + \cdots,  \label{eq6.13}
  \end{eqnarray}
where $\tau = 1 - e_{0}(k_{0}e_{2}+k_{1}e_{3})$ was also defined in (3.10). When the mixing coefficients, $l_{1}$, $j_{1}$, and $j_{2}$ all vanish $\tilde{t}$ reduces to $\tau t$. Notice, however, in contrast to the grand canonical formulation, that for nonzero $l_{1}$ or $j_{1}$ the scaling fields $\tilde{t}$ is now a singular function of $t$ with leading {\em nonlinear} contributions varying as $|t|^{2-\alpha-\beta}$ and $|t|^{2-\alpha}$ ( in place of $t^{2}$, etc.). By the same token, corrections proportional to $\tilde{m}^{2} \sim m^{2}$, arising from the expansion of $F_{\pm}^{\mu}(z)$ and $F_{\pm}^{p}(z)$, will carry the singular factors $|t|^{\gamma-\beta}$ and $|t|^{\gamma}$; moreover, the former actually {\em dominates} the nominally leading term linear in $t$.

For the general canonical order variable, $\tilde{m}$, we find from {\bf I}(2.18)
  \begin{eqnarray}
   \tilde{m} & = & e_{0}m  + e_{0}e_{3}\tilde{s} + (j_{2}+j_{1}k_{1})(e_{0}^{2}/\tau)m^{2} + \cdots, \nonumber \\
  & = & e_{0}m + e_{0}e_{3}Q|\tilde{t}|^{1-\alpha}F_{\pm}^{s}(z;y_{4},\cdots) \nonumber \\
  &   & +\: (j_{2}+j_{1}k_{1})(e_{0}^{2}/\tau) m^{2} + \cdots,  \label{eq6.14}
  \end{eqnarray}
where, from (2.10) for $\tilde{s}$, we find
  \begin{equation}
   F_{\pm}^{s}(z;\cdots) = (2-\alpha)F_{\pm}^{p}(z;\cdots)  - (\beta+\gamma)zF_{\pm}^{\mu}(z;\cdots).  \label{eq6.15}
  \end{equation}
Evidently, $\tilde{m}$ also entails singular terms which, indeed, introduce $|\tilde{t}|^{1-\alpha}$ as a leading correction unless $e_{3}=l_{1}+j_{1}$ vanishes.

Finally, $\tilde{f}(\rho,T)$, the singular part of the Helmholtz free energy, can be expressed as a sum of a scaling piece, which simply extends the original leading form (1.13), plus a series of {\em nonscaling}, singular but higher order corrections arising from field mixing. If we define the scaling functions
  \begin{eqnarray}
  &&  X_{\pm}(z;y_{4},\cdots) = F_{\pm}^{p} - z F_{\pm}^{\mu}, \hspace{0.25in} X_{\pm}^{p} = zF_{\pm}^{p}, \label{eq6.16} \\
  &&  X_{\pm}^{\mu}(z;\cdots) = \pm\; F_{\pm}^{s}(z;\cdots)F_{\pm}^{\mu}(z;\cdots), \hspace{0.1in} X_{\pm}^{s} = \pm\; F_{\pm}^{s}F_{\pm}^{p},  \nonumber
  \end{eqnarray}
the explicit result, recalling (\ref{eq6.8}), is
  \begin{eqnarray}
   \tilde{f}(\rho,T) & = & -Q|\tilde{t}|^{2-\alpha}\mbox{\boldmath\large $[$} X_{\pm}(z;y_{4},y_{5},\cdots) \nonumber \\
  &  & -\: j_{2}QU|\tilde{t}|^{\beta}X_{\pm}^{p}(z;\cdots) \nonumber \\
  &  & \hspace{0.25in}  +\: (e_{0}e_{3}/U)|\tilde{t}|^{1-\alpha-\beta}X_{\pm}^{\mu}(z;\cdots) \nonumber \\
  &  & \hspace{0.25in} +\: j_{2}e_{0}e_{3}Q|\tilde{t}|^{1-\alpha}X_{\pm}^{s}(z;\cdots) + \cdots \mbox{\boldmath\large $]$}.  \label{eq6.17}
  \end{eqnarray}
Evidently the most singular nonscaling correction is of relative order $|\tilde{t}|^{\beta}$ and arises only from the pressure mixing coefficient $j_{2}$ that induces a Yang-Yang anomaly. In as far as this and the other nonscaling corrections are of higher order (in powers of $\tilde{t}$) than the scaling term, they might be regarded as part of a ``singular background piece'', say, $f_{0s}(\rho,T)$. But the singular nature of the canonical scaling fields $\tilde{t}$ and $\tilde{m}$ cannot be so readily sidestepped!

In contemplating these results one may speculate that there might exist better choices of the canonical scaling fields, $\tilde{m}$ and $\tilde{t}$, that would ameliorate the singular mixing terms in (\ref{eq6.13}) and (\ref{eq6.14}) and/or absorb some or all of the nonscaling corrections in (\ref{eq6.17}); however, this seems unlikely to us. Indeed, it is worth recalling that even the concept of a ``nonsingular background'' encounters dangers near criticality in a canonical or Helmholtz formulation. Thus in a symmetric system near $\Tc$ with $m$$\,=\,$$\rho - \rhoc$ one might reasonably expect the background to have the power series expansion
  \begin{equation}
   f_{0}(\rho,T) = f_{\mbox{\scriptsize c}} + f_{1}t + f_{2}m^{2} + f_{1,2}tm^{2} + f_{4}m^{4} + \cdots.  \label{eq6.18}
  \end{equation}
But since the inverse susceptibility $\chi^{-1}(T)$ is given by $(\partial^{2}f/\partial m^{2})_{T}$, the susceptibility itself cannot diverge at $\Tc$ unless $f_{2}$ vanishes identically. Similarly, if $f_{1,2}$ and $f_{4}$ do  not {\em also} vanish one would have $\gamma\leq 1$ and $\delta\leq 3$, both of  which inequalities contradict exact theory and precise experimentation! These observations point, of course, to the fundamental character of a grand canonical or, better, a full field formulation in terms of $p$, $\mu$, and $T$.

\subsection{Finite-size canonical criticality}
\label{sec6.2}

To extend our canonical scaling description to finite systems we may follow Sec.\ II. First, in the set of scaled variables (2.3), we replace $y_{L}=U_{L}\tilde{h}L^{\Delta/\nu}$ by
  \begin{equation}
   z_{L} = B_{L}\tilde{m}L^{\beta/\nu}.  \label{eq6.19}
  \end{equation}
Then, in addition to a nonsingular background free energy $f_{0}(\rho,T;L)$, we may anticipate a singular part, corresponding to (\ref{eq6.17}), of the form
  \begin{eqnarray}
   f_{s}(\rho,T;L) & = & L^{-(2-\alpha)/\nu}\mbox{\boldmath\large $[$} X_{0}(x_{L},z_{L};y_{L4},y_{L5},\cdots) \nonumber \\
  &  & +\: j_{2}L^{-\beta/\nu}X_{1}(x_{L},z_{L};y_{L4},\cdots) \nonumber \\
  &  & + e_{0}e_{3}L^{(1-\Delta)/\nu}X_{2}(x_{L},z_{L};\cdots) \nonumber \\
  &  & +\: j_{2}e_{0}e_{3}L^{(\alpha-1)/\nu}X_{3}(x_{L},z_{L};\cdots) + \cdots \mbox{\boldmath\large $]$}.  \label{eq6.20}
  \end{eqnarray}
Note that the new finite-size scaling functions $X_{0}$ and $X_{3}$ should be symmetric under $z_{L}$$\,\Leftrightarrow\,$$-z_{L}$, $y_{L5}$$\,\Leftrightarrow\,$$-y_{5L}$, etc., while $X_{1}$ and $X_{2}$ are antisymmetric. In the absence of field mixing we recover the obvious finite-size generalization of the scaling form (1.13). However, the pressure mixing coefficient $j_{2}$ generates a {\em non}scaling correction that vanishes as $L^{-\beta/\nu}$ and is antisymmetric in $z_{L}$, $y_{L5}$, etc. The coefficient $l_{1}$, that mixes the chemical potential into the thermal field $\tilde{t}$, produces an antisymmetric correction vanishing as $L^{-(1-\alpha-\beta)/\nu}$.

As in (2.6)-(2.8) we expect that the scaling functions, $X_{j}(x_{L},z_{L};y_{L4},y_{L5},\cdots)$, can be expanded generally in powers of the irrelevant variables, $y_{L4}$, $y_{L5}$, etc., and, also for finite $L$ near criticality, in powers of $x_{L}$ and $z_{L}$, with coefficients $X_{j,kl}^{\mbox{\boldmath\scriptsize $\kappa$}}$ as in (2.8). [See also {\bf K}(4.182)-(4.184).] There is, in fact, a concealed subtlety here: specifically, the particle number $N$ is an integer so that the density $\rho$ (and $m$) are intrinsically discrete variables in a finite system. Away from criticality the free energy surely approaches an analytic function of $\rho$ when $L\rightarrow\infty$; but the degree to which a corresponding smoothness may be assumed in a finite system close to criticality is {\em not} obvious. [Incidentally, the corresponding issue can be raised in connection with the two-Gaussian description of the distribution $P_{L}(\rho;\mu,T)$ in (4.13).] However, in the absence of concrete evidence to the contrary, the assumption that the finite-size canonical free energy, $f(\rho,T;L)$, may be treated as an analytic function through $(\rhoc,\Tc)$ seems highly plausible if used, as here, to determine leading asymptotic behavior when $L\rightarrow\infty$.

Now simulations of simple fluid systems reveal that as, a function of density, $f(\rho,T;L)$ exhibits two peaks for $T\lesssim\Tc$ that correspond to the separation of the two phases. One may then define a finite-size canonical critical point, {\boldmath $($}$\rhoc^{0}(L),\Tc^{0}(L)${\boldmath $)$}, as a point where these two peaks merge. By virtue of the analytic behavior of $f(\rho,T;L)$, such canonical critical points must, in general, be classical in character. However, they will --- at least in simple cases --- approach the bulk critical point $(\rhoc,\Tc)$, whether or not the critical behavior remains classical in the thermodynamic limit. In principle, extrapolating such canonical critical points may help locate the limiting critical point; in practice, however, this has so far proved of limited usefulness \cite{ref7,ref19}: see the numerical behavior revealed in Fig.\ 3 of \cite{ref7} and Fig.\ 1 of \cite{ref19}. Nevertheless, it is of interest to elucidate the asymptotic behavior, especially of $\rhoc^{0}(L)$.

The conditions determining a classical critical point reduce to
  \begin{equation}
   (\partial f_{s}/\partial m)_{T} = 0, \hspace{0.3in} (\partial^{2}f_{s}/\partial m^{2})_{T} = 0.  \label{eq6.21}
  \end{equation}
On expanding the scaling functions in (\ref{eq6.20}) these yield
  \begin{eqnarray}
   0 & = & 2X_{0,02}^{0}z_{L} + j_{2}X_{1,01}^{0}L^{-\beta/\nu} + 2X_{0,02}^{(4)}U_{L4}^{\mbox{\scriptsize c}}L^{-\theta/\nu}z_{L}\nonumber \\
  &  & +\: e_{0}e_{3}X_{2,01}^{0}L^{(1-\Delta)/\nu} + \cdots,  \label{eq6.22} \\
   0 & = & 2X_{0,02}^{0} + 2X_{0,12}^{0}x_{L}+ 6j_{2}X_{1,03}^{0}L^{-\beta/\nu}z_{L}  \nonumber \\
  &  & +\: 2X_{0,12}^{(4)}U_{L4}^{\mbox{\scriptsize c}}L^{-\theta/\nu} + \cdots.  \label{eq6.23}
  \end{eqnarray}
Solving these equations for $x_{L}$ and $z_{L}$ and using (\ref{eq6.13}) and (\ref{eq6.14}) for $\tilde{t}$ and $\tilde{m}$ finally yields the critical temperature as
  \begin{eqnarray}
   t_{\mbox{\scriptsize c}}(L) & = & [\Tc^{0}(L)-\Tc]/\Tc  \nonumber \\
  & = & c_{1}L^{-1/\nu}[1 + c_{2}L^{-\theta/\nu} +j_{2}c_{3}L^{-2\beta/\nu} + \cdots]  \label{eq6.24}
  \end{eqnarray}
and the canonical critical density as
  \begin{eqnarray}
   \rhoc^{0}(L) & = & \rhoc[ 1+ j_{2}b_{1}L^{-2\beta/\nu} + b_{2}L^{-(1-\alpha)/\nu} \nonumber \\
  &  & +\: b_{3}L^{-(2\beta+\theta)/\nu} + \cdots],  \label{eq6.25}
  \end{eqnarray}
where the leading amplitudes are given by
  \begin{equation}
    c_{1} = -X_{0,02}^{0}/D\tau X_{0,12}^{0} \hspace{0.1in} \mbox{and} \hspace{0.1in} b_{1}= -e_{1}X_{1,01}^{0}/2X_{0,02}^{0}.  \label{eq6.26}
  \end{equation}
It is instructive to learn that the asymptotic behavior of $\rhoc^{0}(L)$ has the same form as exhibited by the $k$-loci and the $Q$-loci evaluated at $T=\Tc$: see (3.13) and (4.8). The data for the RPM, however, suggest that the two leading corrections in (\ref{eq6.25}) compete rather strongly so that $\rhoc^{0}(L)$ appears to approach $\rhoc$ nonmonotonically \cite{ref19}.

\section{Conclusion}
\label{sec7}

In this article we have extended to finite systems the ``complete'' scaling theory developed in Part I \cite{ref10} for critical behavior in the thermodynamic limit that incorporates pressure mixing in the scaling fields as well as the irrelevant corrections to scaling. The basic theory is set out in Sec.\ II in a grand canonical or $(p,\mu,T)$ formulation: see (2.1), (2.3), and (1.1)-(1.4). The possibility of finite-size corrections in the scaling fields $\tilde{p}$, $\tilde{h}$, and $\tilde{t}$ [see (1.8)] has been reviewed briefly in Sec.\ II.B and, in Sec.\ II.D, a fairly direct route to detecting such a dependence --- by studying numerically $\mu(\Tc,\rhoc;L)$ --- is proposed.

Section III applied the theory to elucidate the near-critical behavior of the $k$-loci, defined in the $(\rho,T)$ plane by the isothermal maxima of the modified susceptibilities $\chi(T,\rho)/\rho^{k}$: see Fig.\ 1. The usefulness of the $k$-loci in estimating the critical density, $\rhoc$, via simulations is demonstrated for the hard-core square-well fluid in Sec.\ III.B and Figs.\ 2 and 3. It also transpires that the value of $k$ which yields a locus that approaches the critical point ``mostly directly'' provides a reasonable estimate of the Yang-Yang ratio $R_{\mu}$ \cite{ref8,ref9,ref10} that, in turn, provides the most direct measure of the degree to which pressure enters the ordering field $\tilde{h}$. In this way Fig.\ 1(b) provides rather clear evidence of a significant ratio, $R_{\mu}\simeq 0.26$, in the restricted primitive model electrolyte: see Sec.\ III.B.

The behavior of the basic moment ratio $Q_{L}(T;\langle\rho\rangle)$, as defined (following Binder \cite{ref20}) in (1.9), is the topic of Sec.\ IV: see Figs.\ 4 and 5. In particular, the associated $Q$-loci (and $Q^{(k)}$-loci) are determined in Sec.\ IV.A (and IV.B): see (4.8) [and (4.10)] and Fig.\ 6. Of especial interest is the behavior of $Q_{L}(T;\langle\rho\rangle)$ {\em below} $\Tc$, within, up to, and beyond the boundaries, $\rho_{+}(T)$ and $\rho_{-}(T)$, of the two-phase region: see Figs.\ 7-9. For fixed $T<\Tc$ and large enough system sizes, $L$, {\em exact} nontrivial results have been found, as shown in Sec.\ IV.C and D: in particular, the study of the minima in $Q_{L}(T,\langle\rho\rangle)$ [see Fig.\ 9 and (4.38)] lays the foundation for a precise method \cite{ref24} of estimating $[\rho_{+}(T)-\rho_{-}(T)]$ and the coexistence curve diameter $\bar{\rho}(T)$ at higher temperatures very close to $\Tc$.

Of remarkable value for estimating $\Tc$ for asymmetric fluid models is the behavior of $Q_{L}$ evaluated {\em on} the corresponding $Q$-loci: see Fig.\ 10 and the asymptotic expression (5.4) and corresponding plots in Fig.\ 11. Likewise, the estimation of the critical value $Q_{\mbox{\scriptsize c}}\equiv Q_{\infty}(\Tc;\rhoc)$, described in Sec.\ V.B, is important for determining the universality class of criticality. Finally, in Sec.\ V.C and Figs.\ 12 and 13, the estimation of the critical exponent $\nu$ for the highly asymmetric restricted primitive model electrolyte has been described (confirming Ising character).

The issue of a {\em canonical} or $(\rho,T)$ formulation of criticality with corrections to scaling {\em and} pressure mixing is taken up in Sec.\ VI. The basic expression, (\ref{eq6.17}), for the singular part of the Helmholtz free energy is intrinsically more complex than the $(p,\mu,T)$ scaling formulation, entailing an infinite series of ``improperly scaling'' corrections. This formulation provides a basis for determining the asymptotics of the canonical critical points (of classical character) that can be observed in $(N,V,T)$ simulations: see (6.24) and (6.25).

In summary, we believe that the theory developed here and the applications illustrated constitute a solid foundation for future computational studies of criticality that employ systems of finite size.

%\pagebreak

\acknowledgements
The interest of Gerassimos Orkoulas and Erik Luijten and their vital assistance in the computations reported has been much appreciated. The support of the National Science Foundation (through Grant No.\ CHE 99-81772) has been crucial.


\begin{thebibliography}{99}

\bibitem{ref1} See, e.g., as a recent example, G.\ Orkoulas, A.\ Z.\ Panagiotopoulos and M.\ E.\ Fisher, Phys.\ Rev.\ E {\bf 61}, 5930 (2000).
\bibitem{ref2} M.\ E.\ Fisher in ``{\em Critical Phenomena}'', Proceedings of the 51st Enrico Fermi Summer School, Varenna, Italy, edited by M.\ S.\ Green (Academic Press, New York, 1971).
\bibitem{ref3} M.\ E.\ Fisher and M.\ N.\ Barber, Phys.\ Rev.\ Lett.\ {\bf 28}, 1516 (1972).
\bibitem{ref4} J.\ L.\ Cardy, {\em Finite-Size Scaling} (North Holland, Amsterdam, 1988).
\bibitem{ref5} V.\ Privman, {\em Finite Size Scaling and Numerical Simulation of Statistical Systems} (World Scientific, Singapore, 1990).
\bibitem{ref6} J.\ Zinn-Justin, {\em Quantum Field Theory and Critical Phenomena}, 3rd ed. (Clarendon Press, Oxford, 1996) Chap.\ 36.
\bibitem{ref7} See, e.g., G.\ Orkoulas, M.\ E.\ Fisher and A.\ Z.\ Panagiotopoulos, Phys.\ Rev.\ E {\bf 63}, 051507 (2001).
\bibitem{ref8} M.\ E.\ Fisher and G.\ Orkoulas, Phys.\ Rev.\ Lett.\ {\bf 85}, 696 (2000).
\bibitem{ref9} G.\ Orkoulas, M.\ E.\ Fisher and C.\ \"{U}st\"{u}n, J.\ Chem.\ Phys.\ {\bf 113}, 7530 (2000).
\bibitem{ref10} Y.\ C.\ Kim, M.\ E.\ Fisher and G.\ Orkoulas, Phys.\ Rev.\ E (2003) [in press] arXiv:cond-mat/0212145 (6 Dec 2002). This article is here denoted {\bf I} and equations appearing there are labeled {\bf I}(1.1), {\bf I}(3.29), $\cdots$, etc.
\bibitem{ref11} Y.\ C.\ Kim, Ph.D. Thesis, ``Fluid Criticality: Experiment, Scaling and Simulations'', University of Maryland (2002). This work, which contains further details of the analyses presented here, will be denoted {\bf K} and equations therein will be referenced as, e.g., {\bf K}(3.41), etc. Note the remark in \cite{ref28} below concerning normalization of the finite-size scaling formulation.
\bibitem{ref12} J.\ J.\ Rehr and N.\ D.\ Mermin, Phys.\ Rev.\ A {\bf 8}, 472 (1973).
\bibitem{ref13} Y.\ C.\ Kim, M.\ E.\ Fisher and M.\ C.\ Barbosa, J.\ Chem.\ Phys.\ {\bf 115}, 933 (2001).
\bibitem{ref14} V.\ Privman and M.\ E.\ Fisher, J.\ Phys.\ A: Math.\ Gen.\ {\bf 16}, L295 (1983).
\bibitem{ref15} V.\ Privman, Phys.\ Rev.\ B {\bf 38}, 9261 (1988); Physica A {\bf 177}, 241 (1991).
\bibitem{ref16} A.\ D.\ Bruce and N.\ B.\ Wilding, Phys.\ Rev.\ Lett.\ {\bf 68}, 193 (1992).
\bibitem{ref17} N.\ B.\ Wilding and A.\ D.\ Bruce, J.\ Phys.: Condens.\ Matter {\bf 4}, 3087 (1992).
\bibitem{ref18} These questions will be addressed in further detail as part of a critique of the Bruce-Wilding method: Y.\ C.\ Kim and M.\ E.\ Fisher [to be published]; see also {\bf K} Chap.\ 5.
\bibitem{ref19} E.\ Luijten, M.\ E.\ Fisher and A.\ Z.\ Panagiotopoulos, Phys.\ Rev.\ Lett.\ {\bf 88}, 185701 (2002).
\bibitem{ref20} K.\ Binder, Z.\ Phys.\ B {\bf 43}, 119 (1981).
\bibitem{ref21} E.\ Br\'{e}zin and J.\ Zinn-Justin, Nucl.\ Phys.\ B {\bf 257}, 867 (1985).
\bibitem{ref22} H.\ W.\ J.\ Bl\"{o}te, E.\ Luijten and J.\ R.\ Heringa, J.\ Phys.\ A {\bf 28}, 6289 (1995); H.\ W.\ J.\ Bl\"{o}te, L.\ N.\ Shchur and A.\ L.\ Talapov, Int.\ J.\ Mod.\ Phys.\ C {\bf 10}, 1137 (1999).
\bibitem{ref22a} E.\ Luijten, Phys.\ Rev.\ E, {\bf 60}, 7558 (1999).
\bibitem{ref23} M.\ Rovere, D.\ W.\ Heermann and K.\ Binder, J.\ Phys.: Condens.\ Matter {\bf 2}, 7009 (1990).
\bibitem{ref24} Y.\ C.\ Kim, M.\ E.\ Fisher and E.\ Luijten, arXiv:cond-mat/0304032 (1 Apr 2003) [submitted for publication].
\bibitem{ref25} See, e.g. S.-N.\ Lai and M.\ E.\ Fisher, Molec.\ Phys.\ {\bf 88}, 1373 (1996).
\bibitem{ref27} See, e.g., M.\ E.\ Fisher (a) ``Scaling, Universality and Renormalization Group Theory,'' in Lecture Notes in Physics, Vol.\ 186, {\em Critical Phenomena}, edited by F.\ J.\ W.\ Hahne (Springer, Berlin, 1983), p.\ 1-139; (b) Rev.\ Mod.\ Phys.\ {\bf 46}, 597 (1974); (c) Rev.\ Mod.\ Phys.\ {\bf 70}, 653 (1998).
\bibitem{ref28} Note that in {\bf K}, specifically in {\bf K}(4.16), it is tacitly assumed that the scaling function $Y(x_{L},y_{L};y_{L4},\cdots)$ has the dimensions of $L^{(2-\alpha)/\nu}$, while the hyperscaling relation $d\nu = 2-\alpha$ is {\em not} specifically invoked. In the present terms, $Y(x_{L},\cdots)$ and its derivatives, etc., when appearing in {\bf K} should be regarded as measured in units of $\rhoc$. The following corrections should also be noted: (a) in {\bf K}(4.33) a factor $L^{(\gamma-\beta)/\nu}$ should appear in the second term of the second line; (b) in {\bf K}(4.37) a factor $D_{L}/U_{L}$ should appear in the second term; (c) in {\bf K}(4.38) a factor $D_{L}/U_{L}$ is needed on the right hand side; (d) in {\bf K}(4.41) the same factor $D_{L}/U_{L}$ should appear in $a_{\mu}$; (e) in {\bf K}(4.59) the exponent of $L$ in the first term should read $-2\beta/\nu$; (f) in {\bf K}(4.73) the first term should read $[4(Y_{04}^{0})^{2}-5Y_{02}^{0}Y_{06}^{0}]$ while a factor $U_{L}$ is needed in the second term; (g) in {\bf K}(4.75), {\bf K}(4.76), {\bf K}(4.82) and {\bf K}(4.84) the same factor $U_{L}$ should appear in the right hand side; (h) in {\bf K}(4.81) a factor $U_{L}$ should appear in the second term of the left hand side and on the right hand side.

\bibitem{ref29} V.\ Privman and M.\ E.\ Fisher, Phys.\ Rev.\ B {\bf 30}, 322 (1984).
\bibitem{ref30} S.\ Singh and R.\ K.\ Pathria, Phys.\ Rev.\ B {\bf 31}, 4483 (1985); {\em ibid} {\bf 32}, 4618 (1985).
\bibitem{ref31} H.\ Guo and D.\ Jasnow, Phys.\ Rev.\ B {\bf 35}, 1846 (1987).
\bibitem{ref32} A.\ D.\ Bruce, J.\ Phys.\ A: Math.\ Gen.\ {\bf 28}, 3345 (1995).
\bibitem{ref33} E.\ Br\'{e}zin, J.\ Phys.\ (Paris) {\bf 43}, 15 (1982).
\bibitem{ref34} E.\ R.\ Korutcheva and N.\ S.\ Tonchev, J.\ Stat.\ Phys.\ {\bf 62}, 553 (1991).
\bibitem{ref35} X.\ S.\ Chen and V.\ Dohm, Phys.\ Rev.\ E {\bf 66}, 016102 (2002).
\bibitem{ref36} J.\ Rudnick, H.\ Guo and D.\ Jasnow, J.\ Stat.\ Phys.\ {\bf 41}, 353 (1985).
\bibitem{ref37} V.\ Privman and M.\ E.\ Fisher, J.\ Stat.\ Phys.\ {\bf 33}, 385 (1983).
\bibitem{ref38} V.\ Privman and M.\ E.\ Fisher, J.\ Appl.\ Phys.\ {\bf 57}, 3327 (1985); Commun.\ Math.\ Phys.\ {\bf 103}, 527 (1986).
\bibitem{ref39} K.\ Binder, Phys.\ Rev.\ Lett.\ {\bf 47}, 693 (1981).
\bibitem{ref40} K.\ Binder and D.\ P.\ Landau, Phys. Rev.\ B {\bf 30}, 1477 (1984).
\bibitem{ref41} B.\ D\"{u}nweg and D.\ P.\ Landau, Phys.\ Rev.\ B {\bf 48}, 14182 (1993).
\bibitem{ref42} K.\ Binder, Physica A {\bf 319}, 99 (2003).
\bibitem{ref43} C.\ Borgs and S.\ Kappler, Phys.\ Lett.\ A {\bf 171}, 37 (1992).
\bibitem{ref44} A.\ M.\ Ferrenberg and R.\ H.\ Swendsen, Phys.\ Rev.\ Lett.\ {\bf 63}, 1195 (1989).
\bibitem{ref45} J.\ Valleau and G.\ Torrie, J.\ Chem.\ Phys.\ {\bf 108}, 5169 (1998).
\bibitem{ref46} E.\ Luijten, M.\ E.\ Fisher and A.\ Z.\ Panagiotopoulos, J.\ Chem.\ Phys.\ {\bf 114}, 5468 (2001).
\bibitem{ref47} J.\ Valleau and G.\ Torrie, J.\ Chem.\ Phys.\ {\bf 117}, 3305 (2002).
\end{thebibliography}
\end{document}